\documentclass[fleqn,usenatbib]{mnras}

\usepackage{mathptmx}

\usepackage[T1]{fontenc}
\usepackage{ae,aecompl}

\usepackage{graphicx}	
\usepackage{amsmath}	
\usepackage{amssymb}	

\title[XUV observations of PDS 70 ]{X-ray and UV radiation in the planet-forming T-Tauri system PDS 70. Signs of accretion and coronal activity.}

\author[S. R. G. Joyce et al.]{
Simon R. G. Joyce,$^{1}$\thanks{E-mail: sj423@leicester.ac.uk, simonjoyce1701@gmail.com (SRGJ)}
John P. Pye,$^{1}$
Jonathan D. Nichols,$^{1}$
Richard Alexander,$^{1}$\newauthor
Manuel G{\"u}del,$^{2}$
and David Barrado$^{3}$
\\
$^{1}$School of Physics and Astronomy, University of Leicester, University Road, Leicester, LE1 7RH, UK\\
$^{2}$University of Vienna, Dept. of Astrophysics, Turkenschanzstr. 17, 1180 Vienna, Austria\\
$^{3}$Centro de Astrobiologia (INTA-CSIC), ESAC Campus, Camino Bajo del
Castillo s/n, 28692, Villanueva de la Ca$\tilde{n}$ada, Spain
}

\date{Accepted XXX. Received YYY; in original form ZZZ}

\pubyear{2022}

\begin{document}
\label{firstpage}
\pagerange{\pageref{firstpage}--\pageref{lastpage}}
\maketitle

\begin{abstract}
Planet formation takes place in protoplanetary discs around young T-Tauri stars. PDS 70 is one of the first confirmed examples of a system where the planets are currently forming in gaps in the disc, and can be directly imaged. One of the main early influences on planet formation is the lifetime of the protoplanetary disk, which is limited by the intense stellar X-ray and UV radiation. Stellar coronal activity and accretion of material onto the star are both potential sources of XUV radiation. Previous \textit{Swift} observations detected UV emission, which were consistent with a low rate of accretion. We present follow up observations with the XMM-Newton observatory, which observed PDS 70 simultaneously in X-ray and UV in order to determine intensity of XUV radiation in the system, and identify if the source is coronal, accretion, or both. We detect a strong source in both X-ray and UV, with an average X-ray 0.2-12 keV luminosity of $1.37\times10^{30}\ \mathrm{erg\ s}^{-1}$, and a possible flare which increased the luminosity to $2.8\times10^{30}\ \mathrm{erg\ s}^{-1}$.
The UV flux density is in excess of what would be expected from chromospheric emission, and supports the interpretation that PDS 70 has continuing weak accretion less than $\sim10^{-10}\ \mathrm{M_{\odot}\ yr^{-1}}$. The implications of the detected X-ray and UV radiation are that the disc is likely to be in the final stages of dispersal, and will be completely evaporated in the next million years, bringing an end to the primary planet formation process.

\end{abstract}






\section{Introduction}


T-Tauri systems consist of a newly formed star surrounded by a disc of gas and dust, within which planet formation can take place. The transition from T-Tauri system to a star and planets, without a disc, takes approximately 1-10 Myrs \citep{Ribas_2014}. Some of the main processes at work during this period are accretion from the disc onto the star, formation of planets, which can open gaps in the disc, and finally X-ray and UV irradiation of the disc by the central star, which disperses the disc and brings an end to the classical T-Tauri phase. For reviews see e.g. \citealt{Armitage_2018, Ercolano_Pascucci_2017, Gorti_2016, Hartmann_2016, Alexander_2008, Zahnle_Walker_1982}.


PDS 70 provides an opportunity to study this crucial, but short lived, phase in stellar system formation. It is a planet forming system with a protoplanetary disc and 2 confirmed exoplanets around a K7 type star. It is one of the only confirmed examples of observable planets detected in a planet-forming disc. The first detection of a circumplanetary disc around one of PDS 70s planets has also been reported \citep{Christiaens_2019_ApJ_a} which could be the site of moon formation. 


At a distance of 112.39$\pm$0.24 pc (Gaia EDR3) the disc and planets can be directly imaged by ALMA and the detailed structure of the disc is known to have a gap at 17 to 65 AU where the planets are detected (\citealt{Keppler_2018, Keppler_2019}). The disc is at an inclination of 49.7 degrees \citep{Hashimoto_2015} and the star's inclination angle is 50$\pm$8 degrees \citep{Thanathibodee_2020}.

Direct imaging in addition to multi-wavelength observations make it possible to characterise the electromagnetic radiation environment and relate this to the observed structure of the disc. X-ray and UV radiation (XUV) may interact with the gas and dust in the disc through various processes \citep{Alexander_2014}. This radiation is a key driver of disc evolution and evaporation.


The age of PDS 70 has been estimated at 5.4$\pm$1 Myr \citep{Muller_2018}. This is comparable to the approximately 10 Myr lifespan of protoplanetary discs as shown from studies of the occurrence rate of circumstellar discs in young stellar associations between 1 - 100 Myr old \citep{Ribas_2014}. The observed features of the disc, such as large gaps, are also consistent with PDS 70 being a transitional disc. PDS 70 has a highly structured disc, which may be due to a combination of processes including planets clearing gaps, and evaporation of the disc by XUV radiation. Overall, PDS 70 is likely to be in the final stages of planet formation, just before the final dispersion of the disc.


Young low-mass stars are known to emit high levels of XUV radiation from coronal activity. They are still rapidly rotating following their formation, which generates a strong magnetic field and flaring activity (e.g. \citealt{Johnstone_2021}).

A further source of XUV emission in these systems is the accretion shock, where material from the disc flows through accretion columns and impacts onto the star. This heats the plasma to million degree temperatures, causing emission in the XUV range. (e.g. \citealt{Calvet_1998, Hartmann_2016, Hartmann_1994, Koenigl_1991})


PDS70 appears to be at the age where the relative importance of accretion and coronal XUV emission changes as the protoplanetary system evolves. Young T-Tauris have a high accretion rate  (e.g. $\dot{M}_\simeq 10^{-7}$\,M$_{\odot}$\,yr$^{-1}$) of disc material onto the star, which contributes significantly to the UV emission, but may suppress the X-ray emission \citep{Telleschi_2007b}.
As the inner disc is cleared of material, accretion declines, and so does the accretion driven UV emission. In later stages, the coronal emission of the highly active star is dominant, resulting in an increased X-ray luminosity ($\mathrm{L_{x}}$) that is dependent on the stellar spectral type and rotation rate.

The study of XUV radiation from T-Tauri systems therefore provides information on the stellar coronal conditions and activity level, as well as clues to the inner disc structure, matter accretion rate and conditions in the accretion flow (e.g. \citealt{Schneider_2018}).


Variability of the XUV emission of T-Tauri stars has been extensively studied (e.g. \citealt{Venuti_2014, Flaccomio_2018, Cody_2014}). This variability could be due to rotational modulation \citep{Gregory_2006}. As the star rotates, different parts of the surface come in to view. If the accretion stream is confined to columns, as predicted by the magnetospheric accretion model \citep{Koenigl_1991, Bouvier_2007}, rather than accreting relatively uniformly all around the star \citep{Lynden-Bell_1974} this could cause significant variability in the UV emission over the course of PDS 70s $\sim$3 day rotation. But if the accretion flow is stable for years \citep{Venuti_2015}, then the accretion emission variations will be periodic with the stellar rotation.

Variability could also be due to non-uniform distribution of active regions in the stellar corona, similar to the active and quiescent regions seen on the Sun and also inferred from Zeeman-Doppler imaging of the WTTS V410 \citep{Carroll_2012}. If the active regions remain stable for weeks, their signature could be observed as variability in-phase with the rotation rate of the star, primarily in the X-ray emission. However, most flares would not last longer than the 3 days needed for them to be observed in subsequent rotations. Studies of flares on PMS stars show that $\sim$70 per-cent only last $<$1 day, and the longest last $\sim$3.7 days \citep{Wolk_2005, Flaccomio_2018}. An additional complication is that both variation in the accretion flow \citep{Venuti_2015, Robinson_2022}, and coronal flaring activity could vary on time-scales of hours to days, which would produce variability in the XUV emission which is not correlated with the stellar rotation.





The main question for determining the evolutionary state and structure of the inner disc of PDS 70 is whether or not accretion onto the star is still taking place. 
Several studies using different indicators of accretion concluded that PDS 70 is a non-accretor. The NIR imaging is consistent with a large 65 au central cavity \citep{Dong_2012}. \cite{Gregorio_Hetem_2002} classified the system as a non-accreting Weak-line T-Tauri (WTTS) due to the H-$\alpha$ line equivalent width of 2 \AA\, which is much narrower than the 6.6 \AA\ equivalent width threshold for a K7 type star to be an accreting CTTS (see Table.1 of \citealt{Barrado_y_Navscues_2003}). \cite{Long_2018} classified the system as a non-accretor based on lack of Pa-$\beta$ emission. From Swift U-band (3072-3857 \AA) observations \citep{Joyce_2020} it was estimated that the upper-limit of the accretion rate was  $6 \times 10^{-12}$ M$_{\odot}/\mathrm{yr}$ due to lack of excess emission compared to expected chromospheric emission. The re-analysis of this data by \cite{Thanathibodee_2020} found that a higher accretion rate upper-limit of $0.6 - 2.2 \times 10^{-10}$ M$_{\odot}/\mathrm{yr}$ was compatible with the UV data when shorter wavelength filters, which are less affected by photospheric and chromospheric emission,  were considered.


In this paper we present the first observations of PDS 70 with \textit{XMM-Newton}, including simultaneous X-ray and UV data. The much larger effective area of \textit{XMM}, compared to \textit{Swift} makes it possible to detect the X-ray spectrum with a greatly improved signal-noise. The high count-rate also makes it feasible to detect variability in the X-ray spectrum from one exposure to the next.

The observations were planned to study the possible variability due to rotational modulation of the star, and to aid in distinguishing rotational effects from intrinsic variability. This was done by observing PDS70 on 6 occasions spread across two consecutive rotations of the star.

There are several key questions about PDS70 and about protoplanetary systems in general that we plan to investigate. 

First it is necessary to know the source of the XUV radiation, whether it is coronal, accretion shock, or a combination of both. The question of whether accretion is taking place, and if so,  measuring the accretion rate, will help to determine the conditions in the inner disc and the stage of evolution of the system. There are several models of the structure of the accretion region. The data may shed light on the structure and plasma conditions. 

Establishing the quiescent stellar X-ray luminosity, flaring rate and evolution of the X-ray luminosity, are requirements for accurately estimating the evaporation effects on the protoplanetary disc.


In section \ref{sec:Discussion_1_XUV_source} and \ref{sec:Discussion_1_Accretion} the source of the X-ray and UV radiation is discussed to determine if it is coronal or due to accretion.

In section \ref{sec:Discussion_1_Variability} the possibility of variability in the XUV emission due to rotational modulation is investigated. 

In section \ref{sec:Discussion_1_comparison_with_other_T-Tauris_AND_coronal} the XUV emission of PDS 70 is considered in comparison to other T-Tauri stars, both CTTS and WTTS, and also models of stellar coronal evolution, particularly the activity-rotation relation \citep{Wright_2011}.

Finally, the implications of the detected XUV radiation for the evaporation of the disc, and the consequences for the disc dispersal time and planet formation are discussed in section \ref{sec:Discussion_1_disc_evolution}.





\section{Observations}


\begin{table}

	\caption{Table of XMM observations of PDS70. The phase refers to the stellar rotation which has a period of 3.03 days. The first 3 rows are for the 1st rotation of the star, and the last 3 rows are for the 2nd observed rotation of the star. In the remainder of the paper, observations are referred to by their number 1-6 rather than the observation ID ...201-701. The exposure time is for the EPIC PN X-ray camera.}
	\label{Table:Table_1_obs_list}

	\begin{tabular}{c|c|c|c|c|}
	\hline 
	Num & Date & Obs ID & Phase  & exp time\\ 
	& (dd-mm-yyyy) & (0863800...) &    & (s) \\ 
	\hline 
	1 & 15-07-2020 & 201 &  0.0 - 0.06   & 19800\\

	2 & 16-07-2020 & 301 &  0.31 - 0.34   & 13000\\

	3 & 17-07-2020 & 401 &  0.66 - 0.73   & 22500\\

	\hline

	4 & 18-07-2020 & 501 &  0.0 - 0.04   & 14200 \\

	5 & 19-07-2020 & 601 &  0.32 - 0.39   & 22500\\ 
	
	6 & 20-07-2020 & 701 &  0.60 - 0.63   & 13000\\ 
	      
	\hline 
	\end{tabular}

\end{table}



\begin{table}
	\centering
	\caption{Stellar properties of PDS 70.}
	\label{Table:table_2_stellar_properties}

	\begin{tabular}{ccc} 

		\hline
		Parameter & Value & Reference \\
		\hline
		RA (J2000) & 14:08:10.102 & 1\\
		Dec (J2000) & -41:23:53.053 & 1\\
		Spectral type & K7 & 2\\
		Stellar P$_{\mathrm{rot}}$ (days) & 3.03 $\pm$ 0.06 & 6\\
		Age (Myr) & 5.4 $\pm$ 1.0 & 3\\
		Mass (M$_{\odot}$)  & 0.76 $\pm$ 0.02 & 3\\
		Radius (R$_{\odot}$)  & 1.26 $\pm$ 0.15 & 4\\
		Luminosity (L$_{\odot}$)  & 0.35 $\pm$ 0.09 & 4\\
		$T_{\rm eff}$ (K) & 3972 $\pm$ 36 & 4 \\
		Parallax (mas) & 8.8975 $\pm$ 0.01906 & 1\\
		Distance (pc) & 112.39 $\pm$ 0.24 & 1\\
		$M_{\rm disc}$ ($M_{\mathrm{Jup}}$) & 4.5 & 5\\
		$i_{\rm disc}$ (degrees) & 49.7 & 5\\
		\hline
		
	\end{tabular}
	
            1 = Gaia EDR3, Equinox 2000, Epoch J2020, 2 = \cite{Gregorio_Hetem_2002}, 3 = \cite{Muller_2018}, 4 = \cite{Pecaut_Mamajek_2016}, 5 = \cite{Hashimoto_2015}, 6 = \cite{Thanathibodee_2020}

\end{table}




\begin{figure}
	\includegraphics[width=\columnwidth]{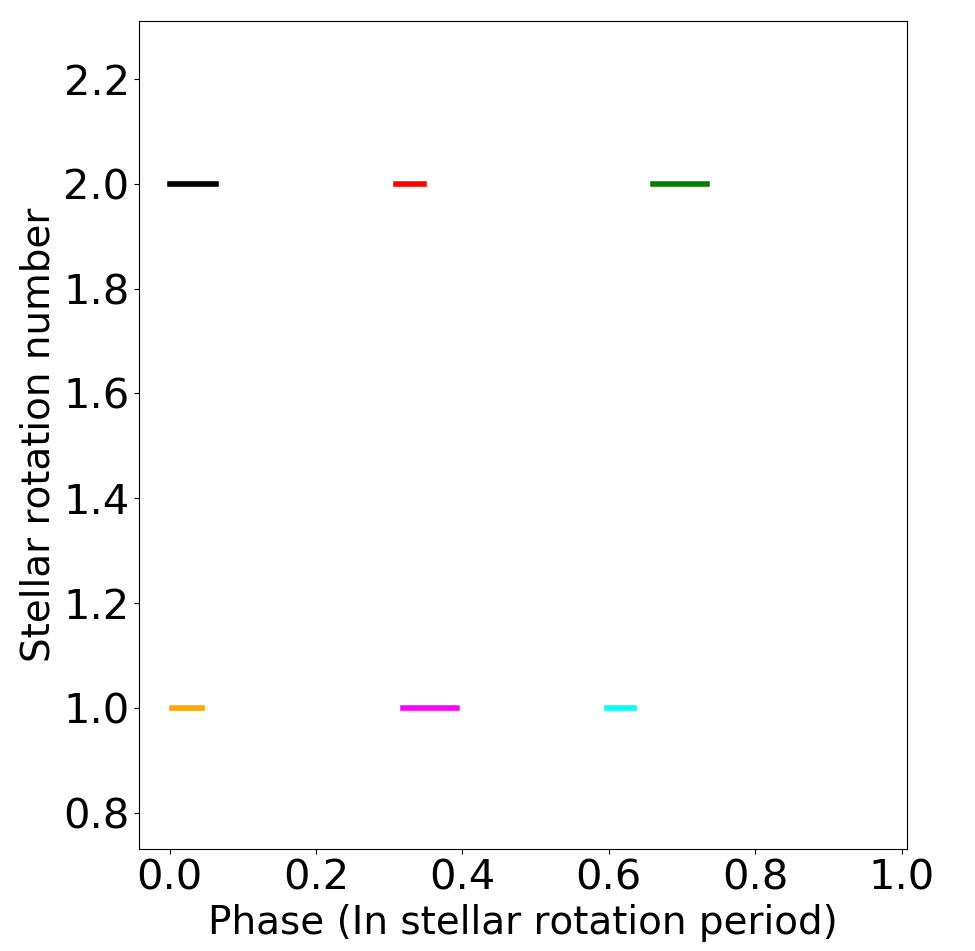}
	\includegraphics[width=\columnwidth]{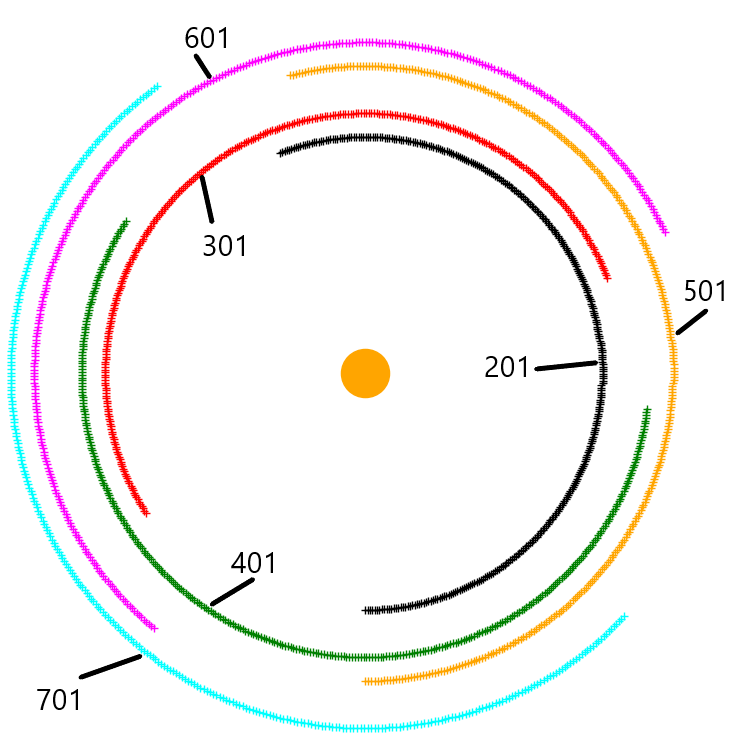}
    \caption{\textbf{Top (A):} Observations of PDS 70 compared to the stellar rotation period. The top row shows the first 3 observations in the 1st stellar rotation, and the second row shows the 3 observations during the second stellar rotation. The lines indicate the duration of the X-ray exposures. UV observations are simultaneous but shorter in duration as several different filters are used. The y axis indicates the 1st and 2nd rotation of the star. \textbf{(B):} Diagram showing the the stellar surface visible during the observations. The lines represent the total circumference of the star seen by each exposure. The same part of the surface is seen for the 1st and 4th observation (black, orange), 2nd and 5th (red, magenta) and 3rd and 6th (green, blue). The annotated numbers refer to the observation ID's in Table \ref{Table:Table_1_obs_list}.}
    \label{fig:Observation_timeline_phase}
\end{figure}


\textit{XMM-Newton} observed PDS 70 six times from 15th to 20th July 2020. The stellar rotation period of PDS 70 is 3.03 days according to analysis of the TESS light-curve \citep{Thanathibodee_2020}. The 6 XMM observations were timed to observe 2 stellar rotations, with 3 observations in each. Individual observations lasted between 12 ks and 21 ks, to give a total exposure time of 98.5 ks, see Table.\ref{Table:Table_1_obs_list}. The observation ID numbers do not start at 101 so in the rest of the paper, the observations are referred to by their observation 'number' 1-6, rather than the observation ID 201 - 701.

The timing of each observation in terms of stellar rotation phase is shown in Fig.\ref{fig:Observation_timeline_phase} A. The observations took place during the same 3 phases of the stellar rotations, so the same regions were observed. Fig.\ref{fig:Observation_timeline_phase} B gives a schematic representation of how much of the stellar circumference was observed during each observation. Each observation detects just over half of the stellar circumference so the timing and spacing of the observations provides total coverage.


During each observation, the X-ray detecting PN, MOS1 and MOS2 cameras \citep{Struder_2001, Turner_2001} were all operating in standard full-frame photon counting mode. The medium filter was used to minimise the contamination from optical, UV and IR light so that the detectors are only exposed to the X-ray emission from the field-of-view. These 3 cameras detect X-rays in the 0.15-15 keV range with a 30 arcmin field of view. They record the time and energy of each detected photon, allowing the data to be viewed as an image, or as a light-curve with timing information, or as a spectrum.\\


The RGS instruments which detect higher resolution X-ray spectra in the 0.35 to 2.5 keV range also operated throughout the observations. However, no significant emission lines were detected.

The Optical Monitor \citep{Mason_2001} was also used to capture UV data. The filter wheel was cycled through the UVW1, UVM2 and UV grism filters to observe using all 3 at least once per observation. Both filters were used with a fast-mode window centred on the target to capture photometry and light-curves in two bands (UVM2 2070-2550 \AA and UVW1 2495-3325 \AA) with exposures lasting $\sim$3 ks. The fast-mode window was operated with the maximum time resolution of 0.5 seconds. Finally, the UV grism was used to capture UV spectra in the 2000-3400 \AA\ wavelength range, with exposures lasting $\sim$4.4 ks per observation.

The data processing was carried out starting from the Observation Data Files (ODF) and following the standard procedures in the data analysis threads available at https://www.cosmos.esa.int/web/xmm-newton/sas-threads . \textsc{sas} version (16.0) was used, with the latest available calibration files. 

For the PN and MOS X-ray data, event files were produced using the \textsc{sas} tasks emproc and epproc respectively. From the event files, images, light-curves and spectra were extracted. By default, spectra are compiled from photons detected during the full exposure. But spectra for specific time intervals during the observations can be extracted by specifying Good Time Intervals (GTI).



\section{Analysis}
\label{sec:Analysis}


\subsection{Source detection}
\label{sec:Analysis_1_source_det} 

\begin{figure}
	\includegraphics[width=\columnwidth]{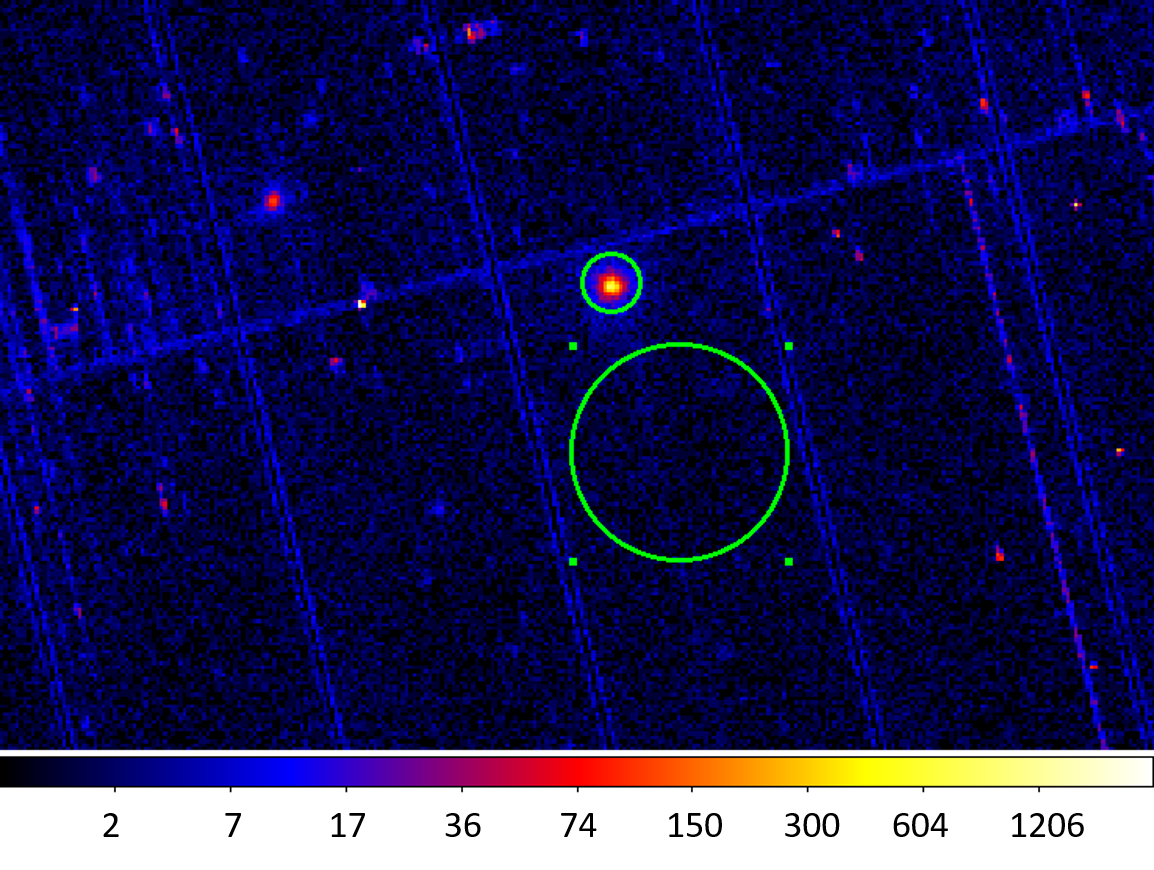}
    \caption{X-ray source detection. The source detection for the 1st (201) observation with the PN X-ray camera. The small circle is the source extraction region, and the larger circle is used to measure the background.}
    \label{fig:Fig_2_XUV_source_detection_image}
\end{figure}

A source at the Gaia DR2 J2020 coordinates of PDS 70 is detected in all of the X-ray and UV observations. Fig.\ref{fig:Fig_2_XUV_source_detection_image} shows the source is clearly detected by the PN camera in the 0.2-12 keV range. Fig.\ref{fig:Fig_Appendix1_XUV_source_detection_image} (A) and (B) show the source detected in the UVW1 2495-3325 \AA\ and UVM2 2070-2550 \AA\ wavelength range respectively. 
The 2nd observation (301) showed a possible transient source close to PDS70, shown in Fig.\ref{fig:Fig_Appendix2_transient_source_image} above and to the left of PDS70. This source did not appear in any of the other observations.

\subsection{Light-curves, X-ray and UV }
\label{sec:Analysis_2_lightcurves} 

\begin{figure}
	\includegraphics[width=\columnwidth]{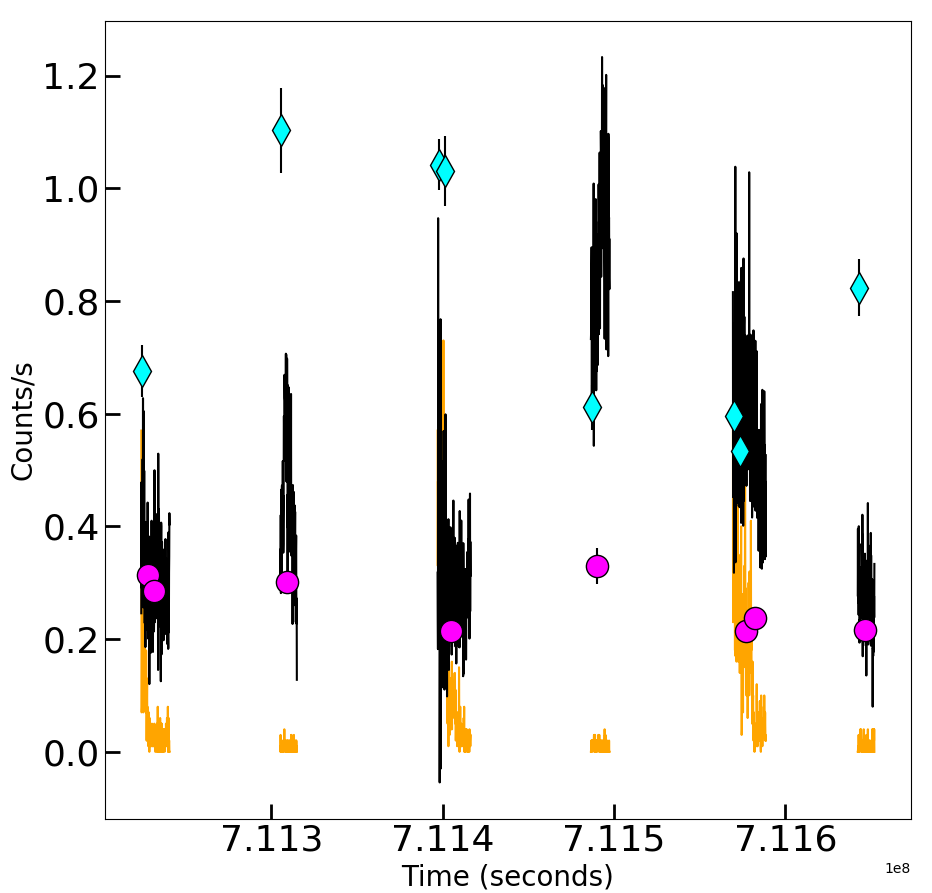}
    \caption{X-ray light-curve for all 6 observations. The source (black) and background (orange) X-ray countrate. The source light-curve is background subtracted, and the background curve is rate-adjusted to the size of the source region. The UV average countrate during each exposure from the UVW1 (blue diamonds) and UVM2 (purple circles) filters are shown for comparison. The UVW1 countrate has been scaled down by -4 counts/s to show on the same scale as the X-ray and UVM2 light-curve.}
    \label{fig:Fig_4_X-ray_lightcurve}
\end{figure}

\begin{figure}
	\includegraphics[width=\columnwidth]{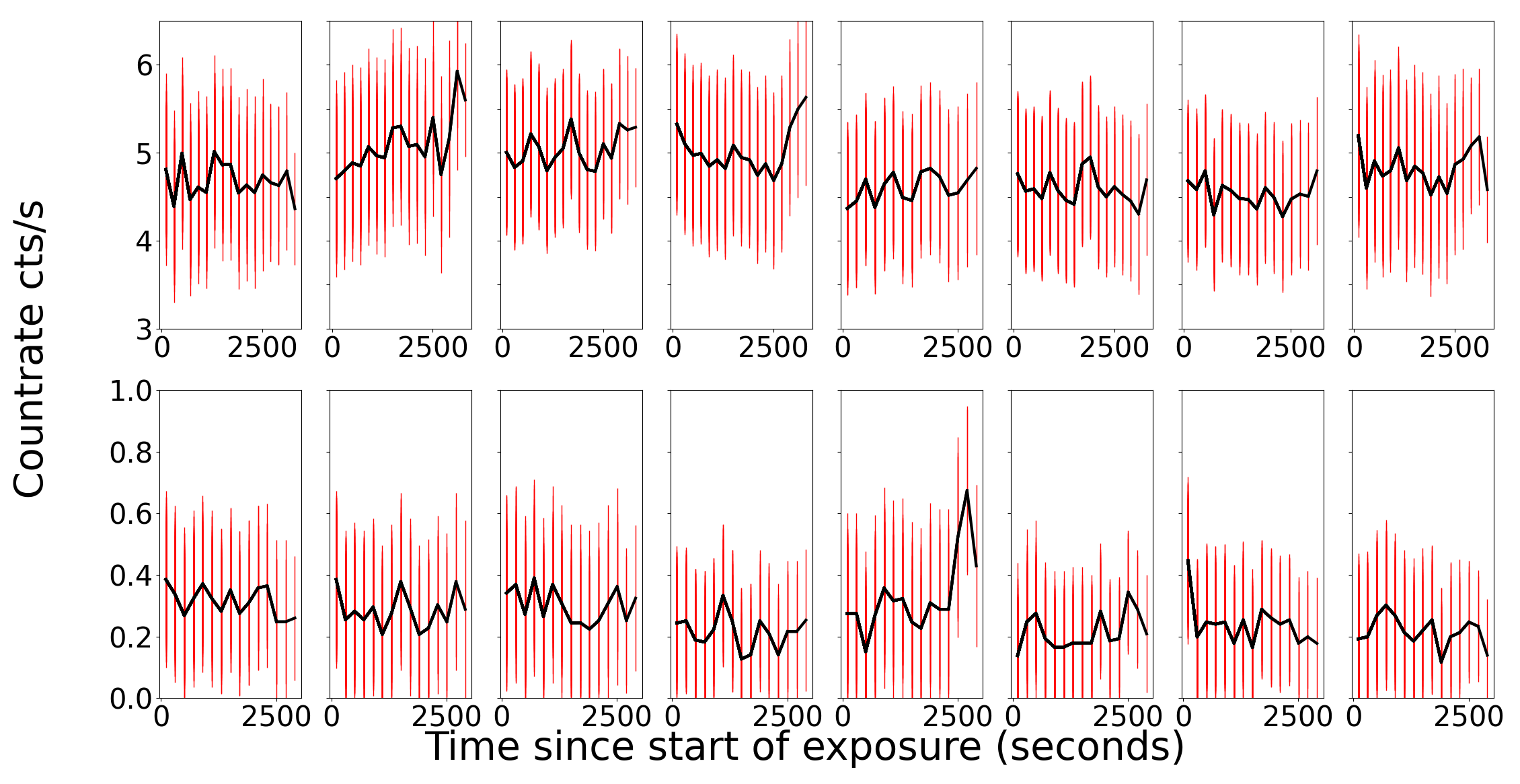}
    \caption{Optical Monitor UV light-curve from UVW1 (Top row) and UVM2 filter (Bottom row). The light-curve is not continuous, each panel corresponds to one exposure, with large intervals in-between as shown in Fig.\ref{fig:Fig_4_X-ray_lightcurve}. There were two exposures during observations 1 and 5, so there are 8 panels for the 6 observations. The possible flare during observation 4 is shown in panel 5, bottom row. The light-curves are binned in 200 second bins for this plot.}
    \label{fig:Fig_5_UV_lightcurve}
\end{figure}

\subsubsection{X-ray light-curves}
\label{sec:X-ray_lightcurves_and_background}

The PN and MOS cameras record timing information about the arrival of photons, so a light-curve can be plotted which shows the count rate during the exposure. 

Light-curves were extracted using the standard procedure 'epiclccorr' tool in \textsc{sas} from the source and background regions. The pointing of the telescope was on target and stable to within 1.43 arcsec throughout the observations, equivalent to 3 pixels in the X-ray images. The plate scale is 0.4765 arcsec/pixel for MOS. 

In Fig.\ref{fig:Fig_4_X-ray_lightcurve} the light-curves show the count rate in the 0.2-10 keV energy range and are binned in 100 second bins to smooth out very short variability. The source light-curve (black) and background light-curve (orange) are plotted together in Fig.\ref{fig:Fig_4_X-ray_lightcurve}. The count rate for the larger background region has been rescaled to be comparable with the size of the source extraction region. 

The orange light-curves show that the majority of the observations took place during periods of very low background, with count-rates of <0.1 $\mathrm{count\ s^{-1}}$. Observations 1, 3 and 5 did have some periods at the start of the exposure where the background count rate was significantly higher, reaching $\sim$ 0.3-0.4 $\mathrm{count\ s^{-1}}$, which is similar to the count rate of the source light-curve (black). The source light-curve has been background subtracted, so the variation in the source count-rate is what remains, independently of the background variation.


The most outstanding feature is that the source count-rate is significantly higher in the 4th and 5th observation, with an average count rate of 0.9$\pm$0.01 and 0.6$\pm$0.01 $\mathrm{count\ s^{-1}}$ respectively, compared to the average count rate of $\sim$0.3$\pm$0.01 $\mathrm{count\ s^{-1}}$ for the rest of the observations. The higher count rate in observation 5 could include some residual contribution from the high background, but the background rate during observation 4 was very low throughout. This shows that the elevated source count rate in these two observations is not an artefact of background subtraction. The uncertainty on the count rate for the 100 second bins is typically 0.05 $\mathrm{count\ s^{-1}}$.

If we take the baseline count rate to be the average of exposures 1,2,3 and 6, then PDS 70 appears to have entered a period of significantly enhanced X-ray emission lasting $\sim$102 ks (28 hours) , assuming the count rate varied smoothly between observations 3 - 6.

\subsubsection{UV light-curves}

The UV count rate from the UVW1 and UVM2 filters is used to see if the UV emission varies in the same way as the X-ray.
The UV data include timing data taken using the fast mode window. The fast mode window is a small area of the image where data is read out every 0.5 seconds so that the count rate variation over time can be measured. For the rest of the image, only the total counts in the full ~3000 sec exposure are known. 
The size of the fast mode window is DX 0.174 arcmin, DY 0.182 arcmin. (10.44 arcsec x 10.92 arcsec).
Fig.\ref{fig:Appen_3_UVW1_show_full_psf_and_FAST_MODE_WINDOW} shows the observation 2 UVW1 image with the fast-mode window highlighted in green. 

One possible cause of variability in the UV count rate could be that the source is not properly centred in the small fast frame window for the full duration of the observation. The pointing is monitored throughout the observation and plotted in the pipeline products. The exposures with the largest variation in pointing are the 2nd and 3rd exposures which have a maximum offset of $\sim$3 pixels. The plate scale is 0.4765 arcsec pixel$^{-1}$, so the 3 pixel pointing variation is well within the 10.44 arcsec window. We conclude that pointing variations would have a negligible effect on the UV countrate , and any variability is attributed to the source.

Fig.\ref{fig:Fig_4_X-ray_lightcurve} and Fig.\ref{fig:Fig_5_UV_lightcurve}, show the UV variability on long and short timescales. Fig.\ref{fig:Fig_4_X-ray_lightcurve} compares the average count rate in each of the $\sim$4000 sec UV exposures to the X-ray count rate. The UVM2 (purple circles) count-rate remains relatively stable between 0.2-0.4 $\mathrm{count\ s^{-1}}$ throughout. In contrast, the UVW1 count rate (blue diamonds) shows significant variation from 4.6 - 5.2 $\mathrm{count\ s^{-1}}$, and is apparently the inverse of the X-ray variability. Note that the UVW1 count rates shown in Fig.\ref{fig:Fig_4_X-ray_lightcurve} have been scaled down by -4 $\mathrm{count\ s^{-1}}$ to allow easier comparison with the X-ray count-rate. 

Fig.\ref{fig:Fig_5_UV_lightcurve} shows the UV light-curve for the UVW1 filter (top) and UVM2 filter (bottom) on shorter time-scales. The count rates are binned in 200 sec bins to smooth out any very short 0.5 sec variability inherent in the observations. There was no evidence of any significant variability on time-scales of $\sim$100's of seconds throughout the UV observations. The UVW1 light-curve does not any short term variability increase during the 4th observation (panel 5 in Fig.\ref{fig:Fig_5_UV_lightcurve}) where the X-ray count rate was enhanced by a factor 5-6. For UVM2, there is a spike in the count rate to 0.6 $\mathrm{count\ s^{-1}}$ lasting 400 seconds, which could indicate a short-lived flare, although the increase is only 1.3 $\sigma$ above the average.


\subsection{UV Grism spectra}
\label{sec:Analysis_UV_Grism} 

Each observation included an exposure with the Optical-Monitor UV grism which can detect UV spectra in the 2000 - 3400 \AA\ wavelength range. This overlaps the ranges covered by the UVW1 and UVM2 filters used for photometry. Exposures lasted just over 3 ks, and observations 1, 3 and 4 included two UV grism exposures. 

The UV spectra from the automatic pipeline processing show prominent emission lines at 2300 and 2600 \AA\ which appear in some exposures but not in others. Inspection of the 2-D images revealed that this was not variability of the source, but was due to the extraction region being placed slightly differently for each exposure, so that different emission lines were included in the extracted spectrum. 

Fig.\ref{fig:Appen_4_Fig_example_UV_Grism_2D} shows an example of the 2-D UV grism dispersion image, with the 1st order extraction region (long red box). The 0th order image of PDS 70 is seen in the smaller red box at the bottom of the image. Further investigation of the 2-D image showed that the emission lines are in fact from contaminating sources, and are not attributed to PDS 70. If they were from PDS 70, a faint vertical spectrum would be visible, as is seen for another source to the right of the image. PDS 70 is visible in the 0th order spectrum because this includes optical light, while the 1st order spectrum only includes UV light.

The non-detection in the UV grism data is consistent with the photometry.
From the instrument handbook, the sensitivity limit of the UV grism is 
$4.6\times10^{-15}\ \mathrm{erg\ s^{-1}\ cm^{-2} \AA ^{-1} }$ for a $1\sigma$ detection after a 5 ks exposure. 
The grism exposures of PDS 70 were $\sim$ 4-5 ks , so the best that could have been achieved was a $\sim 1\sigma$ detection in the UV grism.


\subsection{X-ray spectral fitting}
\label{sec:Analysis_4_X-ray_spec_fit} 

\begin{figure}
	\includegraphics[width=\columnwidth]{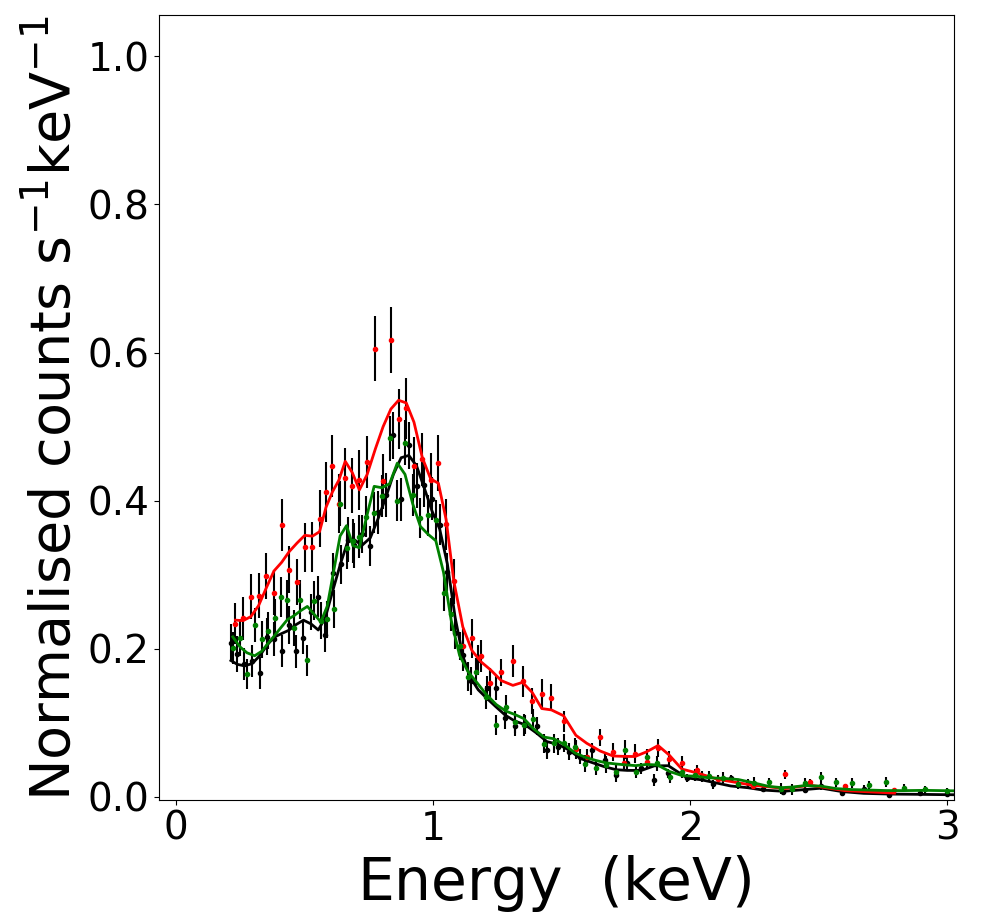}
	\includegraphics[width=\columnwidth]{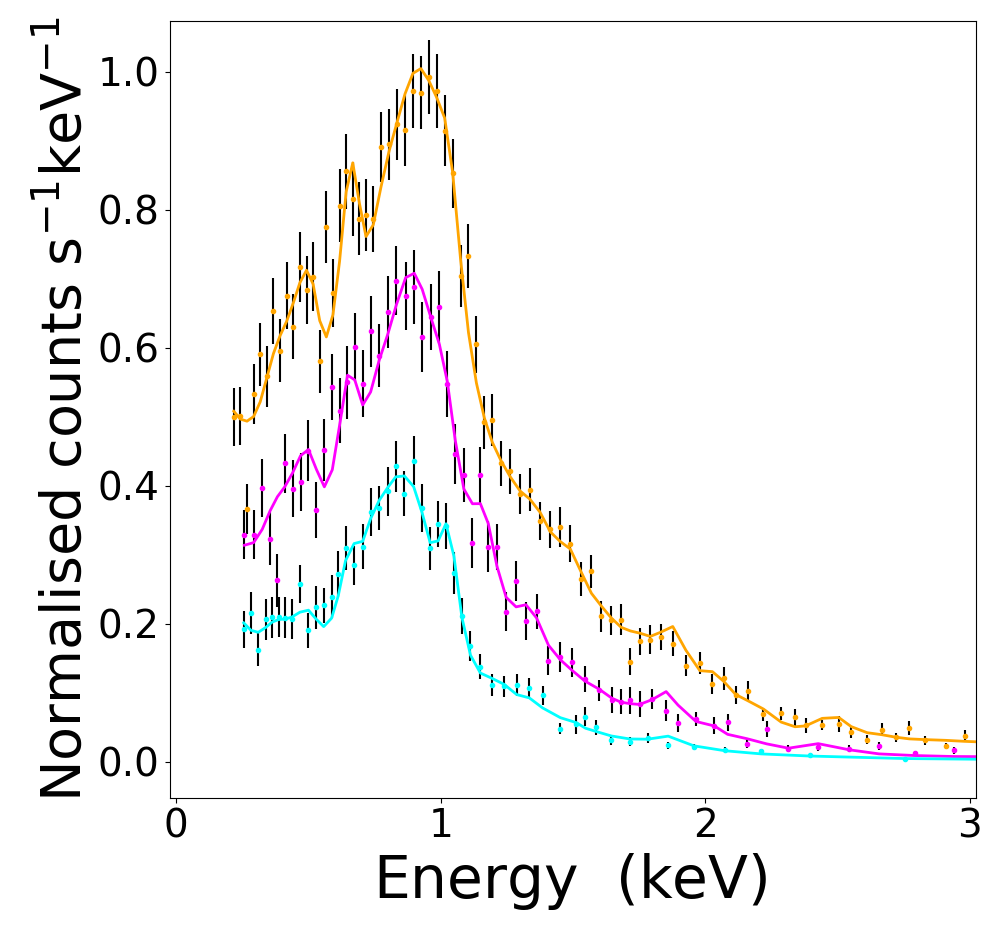}
	
    \caption{X-ray and UV source detection. Top panel (A) X-ray spectral fit for the 3 observations during the 1st stellar rotation. The spectra are fit with a 2-temperature variable abundance optically-thin plasma model shown in black (lower), red (top) and green (middle) for observation 1, 2 and 3 respectively.  (B), X-ray spectral fit for the 3 observations during the 2nd stellar rotation. Orange (top), magenta (middle) and cyan (lower) models correspond to the 4th, 5th and 6th observations respectively. }
    \label{fig:Fig_8_X-ray_spec_1st_rot_A_2nd_rot_B}
\end{figure}

\begin{figure}
	\includegraphics[width=\columnwidth]{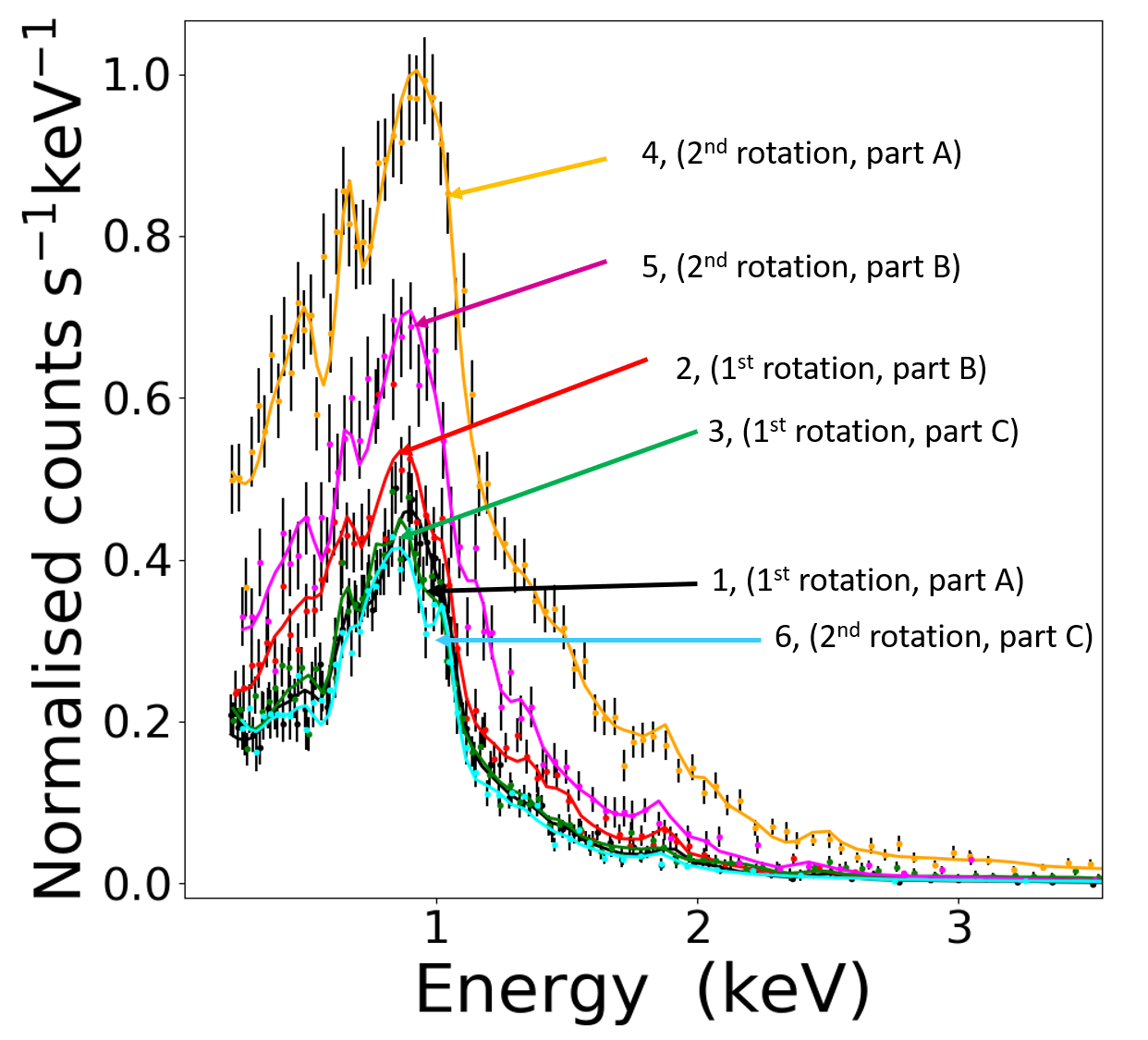}
    \caption{X-ray spectra for all 6 observations fitted with a tbabs*(VAPEC+VAPEC) 2-temperature plasma model. Spectra 1,3 and 6 are almost identical (black, green and cyan) and appear to represent the normal X-ray emission of PDS70. The 2nd (red) spectrum has a slightly increased count rate. The spectrum for the 4th observation has a significantly increased count rate (orange). The 5th observation has an intermediate count rate (magenta). During the final (cyan) observation the spectrum had returned to a state indistinguishable from the 1st and 3rd spectrum. The letters A, B and C in the plot denote observations taken at the same phase in stellar rotation.}
    \label{fig:X-ray_spectra_all_6_QDP_version}
\end{figure}


\begin{table*}
	\centering
	\caption{X-ray spectral fitting results. Flux and luminosity applies to the 0.2 - 12 keV energy range. Upper and lower values are the 68 per cent confidence ranges from \textsc{xspec}. The abundance for Ne, Mg and Fe with respect to solar values is given in the table. The given value applies to both model components as they were linked together. For the rest of the elements, the abundances were fixed at the following values taken from the \citealt{Gudel_2007} survey of T-Tauri stars. He=1, C=0.45, N=0.79, O=0.42, Al=0.5, Si=0.3, S=0.4, Ar=0.55, Ca=0.192, Ni=0.195  The final rows show the UV photometry. Countrates in each filter were converted to flux density using the recommended values of $3.14\times 10^{-16}$ (UVW1) and $1.42\times 10^{-15}$ (UVM2) at https://www.cosmos.esa.int/web/xmm-newton/sas-watchout-uvflux}
	\label{Table:Table_3_spec_fit_results}
	\begin{tabular}{lcccccc} 
		\hline
		\hline
		Parameter &  &  & & & & \\
		\hline
		Observation & 201 & 301  & 401 & 501 & 601 & 701\\
		Rotation & 1 & 1 & 1 & 2 & 2 & 2\\
		
		$N_{\textsc{h}}$ ($10^{22}$ cm$^{-2}$) & 0.02$\pm$0.006   &  0.04$\pm$0.02  & 0.0002$\pm$0.005 & 0.03$\pm$0.003 & 0.03$\pm$0.006 & 0.01$\pm$0.007 \\
		$kT_{1}$ (keV) &  0.21$\pm$0.04  &  0.13$\pm$0.03  &  0.27$\pm$0.01  & 6.9$\pm$3.6 & 0.18$\pm$0.03 & 0.19$\pm$0.06 \\
		$kT_{2}$ (keV) & 0.84$\pm$0.03 &  0.79$\pm$0.03  &  1.03$\pm$0.08   & 0.96$\pm$0.03 & 0.91$\pm$0.04 & 0.77$\pm$0.03\\

		Norm 1 ($\times10^{-4}$) &  0.37$\pm$0.14   & 1.34$\pm$1.39 & 0.007$\pm$0.001 &  1.4$\pm$0.3   & 0.12$\pm$0.03 &  0.15$\pm$0.12 \\
		Norm 2 ($\times10^{-4}$) & 4.06$\pm$0.30 & 6.97$\pm$0.50 &  2.77$\pm$0.36 &  13.9$\pm$0.8   & 9.0$\pm$0.54 &  3.82$\pm$0.31 \\
		
		Ne & 0.48$\pm$0.17  & 0.44$\pm$0.12  & 0.78$\pm$0.46  & 0.23$\pm$0.17 & 0.17$\pm$0.2 & 0.69$\pm$0.17 \\
		Mg & 0.15$\pm$0.09  & 0.27$\pm$0.09  & 0.05$\pm$0.19  & 0.13$\pm$0.07 & 0.04$\pm$0.1 & 0.15$\pm$0.09 \\
		Fe & 0.12$\pm$0.01  & 0.08$\pm$0.01  & 0.12$\pm$0.02  & 0.07$\pm$0.01 & 0.08$\pm$0.01 & 0.08$\pm$0.01 \\

		$\chi^{2}_{\mathrm{red}}$ & 1.02 & 1.48 & 1.035 & 1.14 & 1.56 & 1.27\\
		
		\hline
		
		$F_{\mathrm{X, abs}}$ (erg cm$^{-2}$ s$^{-1}$)($\times10^{-13}$) & 4.21$\pm$0.1   & 5.85$\pm$0.3 & 6.6$\pm$0.4  & 16.2$\pm$0.9 & 7.6$\pm$0.2 & 3.9$\pm$0.1 \\	
		
		$F_{\mathrm{X, unabsorbed}}$ (erg cm$^{-2}$ s$^{-1}$) ($\times10^{-13}$) &  5.02  & 8.28 & 6.63  & 18.70 & 9.21 & 6.75\\	
		
		$L_{\mathrm{X, unabsorbed}}$ (erg s$^{-1}$) ($\times10^{30}$) &  0.76 $\pm$ 0.04  & 1.25 $\pm$ 0.06 &   1.00 $\pm$ 0.04 & 2.82 $\pm$ 0.20 & 1.39 $\pm$ 0.13 & 1.02 $\pm$ 0.11\\	
		
		$L_{\mathrm{X}}/L_{\mathrm{bol}}$  & -3.2   & -3.0 & -3.1  & -2.7 & -3.0 & -3.1 \\	
		
		\hline
		
		$\mathrm{Average} \ F_{\mathrm{X, unabsorbed}}$ (erg cm$^{-2}$ s$^{-1}$) ($\times10^{-13}$) &    &  & 9.09  &&&\\
		$\mathrm{Average} \ L_{\mathrm{X, unabsorbed}} $ (erg s$^{-1}$) ($\times10^{30}$) &    &  & 1.37  &&&\\
		
		\hline
		UV Flux density (erg s$^{-1}$ cm$^{-2}$ \AA$^{-1}$) &&&&&&\\
		UVW1(2495 - 3325 \AA) ($\times10^{-15}$)  &  1.47$\pm$0.01  &  1.60$\pm$0.02  &  1.58$\pm$0.01  &  1.45$\pm$0.01  &  1.44$\pm$0.01  &  1.51$\pm$0.02  \\
		  & - & - & 1.58$\pm$0.02 & - & 1.42$\pm$0.01 & - \\
		UVM2(2070 - 2550 \AA) ($\times10^{-16}$)  &  4.45$\pm$0.2  &  4.3$\pm$0.2  &  3.05$\pm$0.2  &  4.7$\pm$0.5  &  3.05$\pm$0.2  &  3.07$\pm$0.2  \\
		  & 4.05$\pm$0.2 & - & - & - & 3.38$\pm$0.2 & - \\
		
		\hline
	\end{tabular}
\end{table*}


 

 

 

 

 

%


\subsubsection{Spectral extraction regions and time intervals}

Spectra were extracted from the PN and MOS event files using the \textsc{sas} task 'evselect'. A circular region in the image for extracting the source spectrum was defined by centring the circle at the Gaia DR2 coordinates for PDS70 (epoch J2020). A circle of radius 30 arcsec is used to capture the full extent of the source in Fig.\ref{fig:Fig_2_XUV_source_detection_image} . A slightly smaller 25 arcsec circle was used for the 2nd observation (0863800301) to avoid including the nearby transient source in Fig.\ref{fig:Fig_Appendix2_transient_source_image}. Background spectra are extracted from a larger $\sim$100 arcsec circle on the same CCD chip in a region with no visible sources. The background spectrum is scaled to account for the difference in the size of the extraction region. Finally 'arfgen' and 'rmfgen' are used to generate the ancillary files required to calibrate the spectrum when it is loaded in \textsc{xspec} for model fitting.

The background subtraction carried out when extracting the spectra should remove the effect that periods of high background have on the final source spectrum. To make sure there is no residual effect, spectra were extracted for specific time periods identified as low background according to the background count-rate threshold of <0.1 $\mathrm{count\ s^{-1}}$. These are refereed to as Good Time Intervals (GTIs) in the \textit{XMM} documentation.

The spectra from just the low-background time intervals, compared to the spectra from the full exposure showed no significant difference for the 1st and 3rd observations.
 The period of high background was relatively short for both of these exposures, lasting 20 and 36 per cent of the exposure time respectively. The spectrum extracted from the high-background time interval does show excess scatter at around 1 keV and below 0.5 keV, but there are low numbers of counts in these bins. The excess scatter is averaged out when the spectrum is extracted from the full exposure. 
 
The 5th observation was affected by high background for 59 per cent of the observation. The spectrum from the full exposure does show an excess of counts at  $\sim$1 keV which is not seen in the spectrum from the low background time interval. Therefore, in all of the subsequent spectral analysis, spectra for the 5th observation are extracted from the low background time interval only. The overall exposure time for the 5th observation was longer than for most of the exposures (22.5 ks), so the remaining GTI exposure time is 7.9 ks seconds which still produces a high quality spectrum.

Spectra from the PN and MOS1/2 cameras can be combined because all 3 cameras operate simultaneously. The combined spectra were produced using the 'epicspeccombine' tool. The spectra were binned with a minimum of 25 counts per bin. For the 1st exposure, this resulted in 77 bins, compared to 67 from the PN camera alone. 

For X-ray spectral analysis, the \textsc{xspec} software was used with APEC and VAPEC \citep{Smith_2001} plasma models. X-ray spectra require individual response files which were generated with the \textsc{sas} tasks rmfgen and arfgen. 

A comparison of the 6 spectra is shown in Fig.\ref{fig:Fig_8_X-ray_spec_1st_rot_A_2nd_rot_B} panel A and Fig.\ref{fig:Fig_8_X-ray_spec_1st_rot_A_2nd_rot_B} panel B which are for the 1st and 2nd rotation of the star. The phases are 0 (black, orange), 0.3 (red, magenta) and 0.6 (green, cyan) to compare spectra from the same phase of stellar rotation. A comparison of all 6 spectra together is shown in Fig.\ref{fig:X-ray_spectra_all_6_QDP_version}.

Spectra from the 1st, 2nd, 3rd and 6th observation are very similar. The 4th spectrum (orange) is significantly different, with more than double the count rate across the 0.2 - 2.0 keV range. The 5th spectrum (magenta) is higher count rate than the average, but not as much as the 4th spectrum.

The variation seen in the X-ray light curves is seen again in the spectra. This indicates that the X-ray source was constant for rotation 1, regardless of which part of the star was observed, as shown by the similarity at all phases in rotation 1. This is consistent with a model where the X-ray source is uniform across the stellar surface, or is independent of the stellar surface/rotation. 
In the second rotation, the 1st phase spectrum has a significantly higher count rate than all 3 of the rotation 1 spectra. This indicates a change in the X-ray source. The spectrum from rotation 2: phase 2 has a lower count rate than rotation 2: phase 1, but still higher than the average from rotation 1. The 3rd phase of rotation 2 shows a spectrum very similar to the rotation 1 observations, indicating that the source has returned to the baseline as defined by rotation 1.

\subsubsection{Spectral Fitting}
Fitting the spectra with a 2 temperature APEC plasma model achieved a statistically acceptable fit with $\mathrm{\chi^{2}_{red}}$ from 1.3 - 2. However, the difference in elemental abundance between the two temperature components was unrealistically large, 2.67 in the cool component and 0.136 in the hot component, where solar abundance is 1. 

The APEC model sets the abundance for all elements to the same value, so a VAPEC model was investigated which allows abundances for each element to be set individually. It was noted that when the abundances are fixed at 1 (solar abundance), the fit is generally poor at 0.7, 1.0 and 1.5 keV. Varying the abundance of each element in turn showed that these energy ranges are most affected by the Ne, Mg and Fe abundance. 
Several different settings for the abundances were tested based on analysis of other T-Tauri stars found in the literature. The following 3 methods were tested on the 201 spectrum for PDS70 to provide a consistent comparison.

The abundances listed in \cite{Gudel_2007} (Footnote on page 364) derived from fitting spectra for a large number of T-Tauri stars in the Taurus molecular cloud survey produced a good fit with $\mathrm{\chi^{2}_{red}}$ 1.24. These abundances are listed in the caption of Table.\ref{Table:Table_3_spec_fit_results} of this paper.
A slightly different approach was also tested following the method of \cite{Schneider_2018} used to fit the X-ray spectrum of T-Tauri itself. In this method the elements are put into 3 groups according to their first ionisation potential (FIP). Abundances are set at 0.82 for high FIP elements, and 0.35 and 0.32 for mid and low FIP elements. This method produced a slightly worse fit when applied to PDS 70 with $\mathrm{\chi^{2}_{red}}$ 1.69.

\cite{Skinner_2017} fit an X-ray spectrum of Lk Ca 15 which has a similar age and spectral type to PDS 70 (K5 and K7 respectively). The method was to set all abundances to 1, but allow Ne and Fe to vary. Again, this produced a reasonable fit for PDS 70, with $\mathrm{\chi^{2}_{red}}$ 1.49. 

The method which was found to provide the best fit for PDS 70 is a hybrid of the previous methods, whereby the abundances are all set to the values found in \cite{Gudel_2007}, but the Ne, Mg and Fe are allowed to vary, similar to the \cite{Skinner_2017} method. The Ne, Mg and Fe abundances were linked between the two temperature components of the model so that they had to vary in unison. It was found that allowing them to vary independently produced a very good fit statistically, but required widely differing abundances between the two components. 
When the abundances in the two VAPEC components were linked, statistically acceptable fits were found for all 6 spectra with reasonable abundance values in comparison to previous studies of other T-Tauris. The fits using this method were an improvement over all the other methods tried for PDS 70, with a $\mathrm{\chi^{2}_{red}}$ of 1.02 for the 201 spectrum. The $\mathrm{\chi^{2}_{red}}$ values are listed in Table.\ref{Table:Table_3_spec_fit_results} along with the fit parameters for each of the 6 spectra.
For spectrum 301, the fit to the full spectrum was relatively poor, with $\mathrm{\chi^{2}_{red}}$ 2.48. Inspecting the residuals showed that most of the spectrum was a good fit, except above 3 keV where there were very few counts in the spectrum. Limiting the fit to the 0.2-3.0 keV range resulted in a fit of $\mathrm{\chi^{2}_{red}}$ 1.47

It is noticeable that there is a significant decline in the Ne abundance during the 4th and 5th observations when the X-ray flux was enhanced. This could be due to material from lower in the stars atmosphere, where the abundances are different, being brought up during flare activity. However, this would contradict studies of the Sun \citep{Feldman_1992} which have shown that Ne is more abundant in the photosphere than the corona, and is enhanced during flares. Further study with high resolution UV spectra will be needed to determine the element abundances more conclusively.

The X-ray light-curve shows that the count-rate during the 4th observation was more than double the count rate of the other observations. The spectrum also showed significant differences during this period. The 1st, 3rd and 6th spectra are very similar, and represent a baseline for comparison. They have a $\sim$0.2 keV component and a $\sim$0.8 keV component. Fitting the 501 spectrum resulted in a much higher temperature of 6.6 $+/-$3 keV for one component, while the 0.96 keV temperature component was not significantly changed. The normalisation of both components was free to vary, and norm 2 showed an increase by $\times 2.6$ compared to the average of norm 2 in the other spectra. There was no change in the ratio of the normalisation between the two model components. This indicates that the increase in flux was caused by an increased Emission Measure from all the plasma components, but the change in the spectral shape was driven by an increase in temperature, rather than a change in the relative contributions of each component.

The X-ray flux in the 0.2-12 keV energy range is derived from the best fit model and listed in Table.\ref{Table:Table_3_spec_fit_results}. The unabsorbed flux is calculated by setting the $N_{H}$ absorption model component to zero. The luminosity is calculated from the unabsorbed flux so that it represents the intrinsic luminosity of the source without the absorption effect of intervening material in the interstellar medium. The distance used is 112.39 pc (Gaia EDR3). The average luminosity for the 4 quiescent observations is 1.0$\pm$0.1$\times10^{30}$ erg s$^{-1}$, while the luminosity increases considerably for the 4th observation to 2.82$\pm$0.02$\times10^{30}$ erg s$^{-1}$.





\section{Discussion}

\subsection{Source of X-ray and UV emission}
\label{sec:Discussion_1_XUV_source} 

There are two main potential sources of XUV emission in the PDS 70 system. 
Observations of active stars show that coronal magnetic activity heats plasma to temperatures of 1-100 MK \cite{Gudel_2004}, with an average for T-Tauris of 10 - 20 MK \cite{Telleschi_2007a}. As a young, $\sim$ 5 Myr old star, PDS 70 is rapidly rotating, which is likely to generate saturated levels of magnetic activity and associated X-ray emission. The coronal activity produces UV emission, but the observed spectrum of the K7 type main-sequence star HR8086 (61 Cyg B), which is of similar spectral type as PDS 70, shows that there is only low level UV emission at wavelengths between 2000 to 3000 \AA. 

The other potential source of XUV emission is accretion heated plasma, which flows from the inner accretion disk, and impacts on the star, producing hot-spots with temperatures of 1-2 MK. The expected temperature of the accretion hot-spots can be calculated based on the stellar mass and radius using equations 2 and 9 from \cite{Calvet_1998} for the free-fall velocity $v_{s}$ and shock temperature $T_{s}$. 

\begin{equation}
\label{eqn:shock_velocity}
v_{s} = 307\times \left( \frac{M}{0.5M_{\odot}} \right)^{1/2} \left(\frac{R}{2R_{\odot}} \right)^{-1/2}\left(1- \frac{R}{R_{t}}\right)^{1/2}\ (\mathrm{km\ s}^{-1})
\end{equation}

\begin{equation}
\label{eqn:shock_temperature}
T_{s} = 8.6\times10^5 \times \left(\frac{M}{0.5M_{\odot}}\right) \left(\frac{R}{2R_{\odot}}\right)^{-1}\ (\mathrm{Kelvin})
\end{equation}

For the mass and radius of PDS 70, $M$ = 0.76 M$_{\odot}$ and $R$  = 1.26 R$_{\odot}$ from Table.\ref{Table:table_2_stellar_properties} and using a disc truncation radius of $R_{t} = 5\times R_{*}$ as used in \cite{Calvet_1998}, this results in a free-fall velocity of 430$\pm$30 km/s and a shock temperature of 2.1$\pm$0.3 MK. This temperature of plasma would produce emission peaking in the UV (2000-4000 \AA) (See \citealt{Calvet_1998}, Fig 4) and soft X-ray $\sim$0.6 keV range \citep{Robrade_2014}.

Apart from the plasma temperature, accretion and coronal plasma differ in electron density, with coronal plasma having a low density of $n_{\mathrm{e}}  <  3\times10^{10}\  \mathrm{cm}^{-3}$, whereas accretion results in higher plasma densities of $n_{\mathrm{e}} > 10^{11} \  \mathrm{cm}^{-3}$ \citep{Ness_2004}.

If accretion were a major source of X-ray emission in the PDS 70 system, the cool $\sim$0.2 keV ($\sim$2 MK) component would be expected to have an emission measure (EM) comparable to the EM of the hot >0.8 keV ($\sim$9 MK) component, based on the normalisation values from the spectral fitting in Table.\ref{Table:Table_3_spec_fit_results} This is not observed. Instead, fitting the quiescent spectra shows that >90 per cent of the EM is from the hot >0.8 MK component, except for the 2nd observation where the cool component increased to 20 per cent. This supports a coronal origin for the majority of the X-ray emission. 

From the low resolution EPIC X-ray spectra, it is not possible to measure the electron density, except by using the EM and making an assumption about the size of the emitting region. Other studies have used the XMM RGS high resolution to resolve the O\textsc{vii} line and obtain the electron density from the line ratios. The RGS observations of PDS 70 did not result in line detections. 

The UV photometry of PDS 70 indicates a higher level of flux than would be detected from coronal emission alone, if accretion were not occurring. Both the X-ray spectra and UV photometry therefore support a scenario where the majority of X-ray flux is emitted by the magnetically active stellar corona, while a low level of accretion is producing UV emission, in excess of the expected coronal and chromospheric emission.



\subsection{Variability of XUV emission}
\label{sec:Discussion_1_Variability} 

A comparison of the variability seen in the X-ray and UV countrate can indicate if the emission is likely to be from the same emission region and influenced by the same processes such as coronal flaring or variability in the accretion rate. If the emissions at different regions do not vary in the same way, this may suggest that they are driven by independent processes.

\begin{figure}
	\includegraphics[width=\columnwidth]{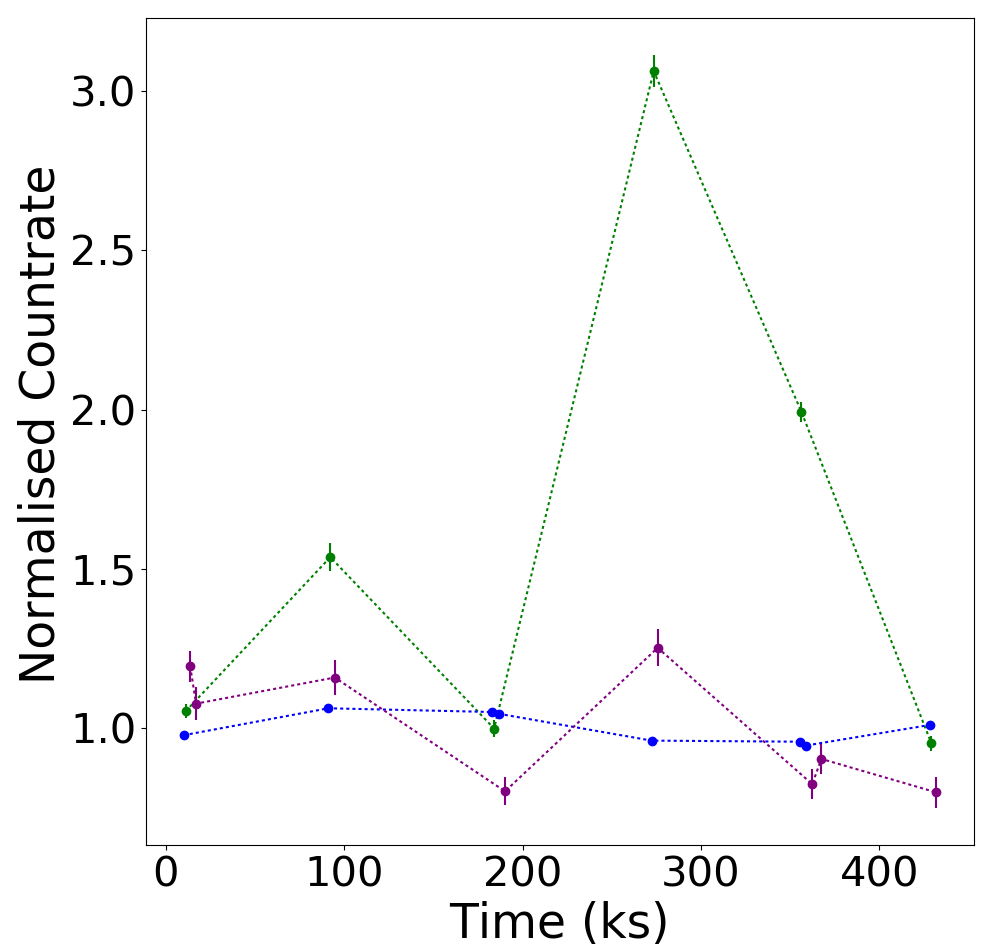}
    \caption{Variability of the X-ray and UV flux over the 6 observations of PDS 70 viewing 2 stellar rotations. The X-ray (green), short wavelegnth UV: UVM2 (purple) (2070-2550 Å) and long wavelength UV: UVW1 (blue) (2495-3325 Å) lightcurve after binning so that each exposure has one datapoint. }
    \label{fig:X-ray_and_UV_variability_obs_by_obs}
\end{figure}

Fig.\ref{fig:X-ray_and_UV_variability_obs_by_obs} shows the X-ray (green), UVM2 (purple) (2070-2550 \AA) and UVW1 (blue) (2495-3325 \AA) lightcurve after binning so that each exposure has one datapoint. The countrate has been normalised to the average of the 6 observations so that the UV can be compared to the X-ray more easily. 
The dashed lines are to guide the eye so that the order of the data points can be easily seen. The dashed lines do not indicate interpolation of the countrate between the observations. 

It can be seen that the X-ray shows a slight positive trend, mostly due to the exceptionally high countrate in the 4th exposure. The UV lightcurves on the other hand both show a slight decline in countrate over the 6 exposures. In all cases, the long term trends are not significant, and are dominated by the point-to-point variability.
To determine if the variability in each lightcurve is statistically significant, the degree of variation from the average is calculated in terms of the 1 sigma error bars. 
The X-ray lightcurve shows clear variability in the 4th exposure, with a sigma of 28.

The UV lightcurves show only a small amount of variability, which is only marginally significant at 5 sigma for the UVM2 (purple) lightcurve. The UVM2 lightcurve does show its maximum variability at the same time as the X-ray source during the 5th exposure. The X-ray and UVM2 lightcurves also show the same pattern of rise and decline in countrate for each observation, although the variation is less extreme in the UVM2 lightcurve than in the X-ray. This could indicate that the variability in both of these wavelength ranges is driven by the same process and emission region. It is expected that the amplitude of the UV variability would be less than for the X-ray because of the greater contrast at shorter wavelengths.
The UVW1 (blue) lightcurve is generally consistent with no variability for the last 5 observations, although the first observation does show a 6 sigma variation when compared to a line of best fit. Fitting a line to the UVW1 data showed a slight trend of decreasing count rate over the 6 observations.  

A possible scenario to interpret these observations is that the longer wavelength (UVW1) source is less variable because the flares produced by magnetic coronal activity are not as bright at longer wavelengths. If the UVW1 wavelength range is dominated by accretion emission, this shows that the accretion rate and accretion column structures are fairly stable over the 6 days observed, and are uniformly distributed around the star, as inferred from the lack of rotational modulation in the UVW1 lightcurve. It could also be that the UVW1 includes a relatively large proportion of chromospheric emission which would hardly be variable. It has been shown that as accretion rates decline, the remaining UV emission is dominated by the chromospheric contribution \citep{Ingleby_2011}.

The shorter wavelength (UVM2) lightcurve shows more variability and is varying in a similar way to the X-ray, as this wavelength range is dominated by the highly variable hot coronal flares. The large change in countrate in the 4th observation compared to the 1st observation, which are both at the same phase in stellar rotation,  shows that this event is due to a real change in the coronal emission rather than rotational modulation alone.

%

\subsubsection{Correlations between X-ray and UV variability}

In this section we study the possible correlations between X-ray and UV emissions on shorter timescales of 100's of seconds. The following analysis is based on that carried out for T-Tauri by \cite{Schneider_2018}. For T-Tauri, it was found that there was a possible anti-correlation between the X-ray and UVW1 countrate. In that case it was proposed that increased accretion, as indicated by enhanced UVW1 emission, somehow suppresses coronal activity and the associated X-ray emission.

\begin{figure}
	\includegraphics[width=\columnwidth, height=0.8\columnwidth]{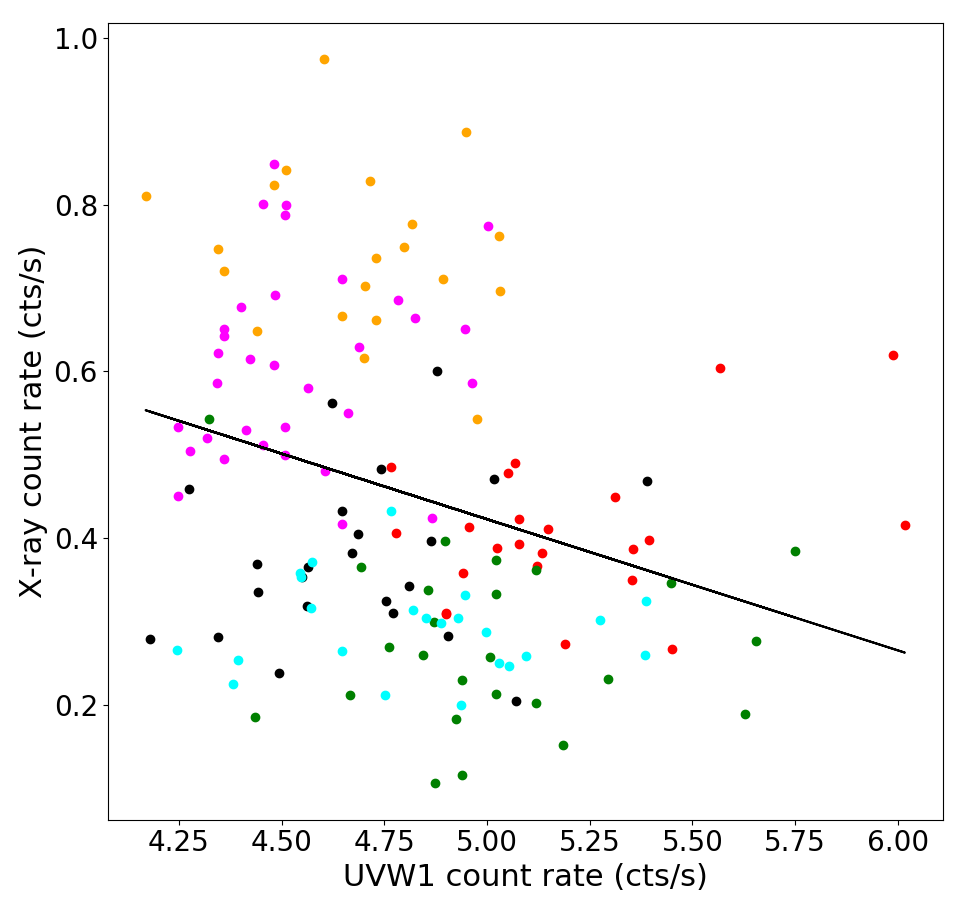}
	\includegraphics[width=0.98\columnwidth, height=0.8\columnwidth]{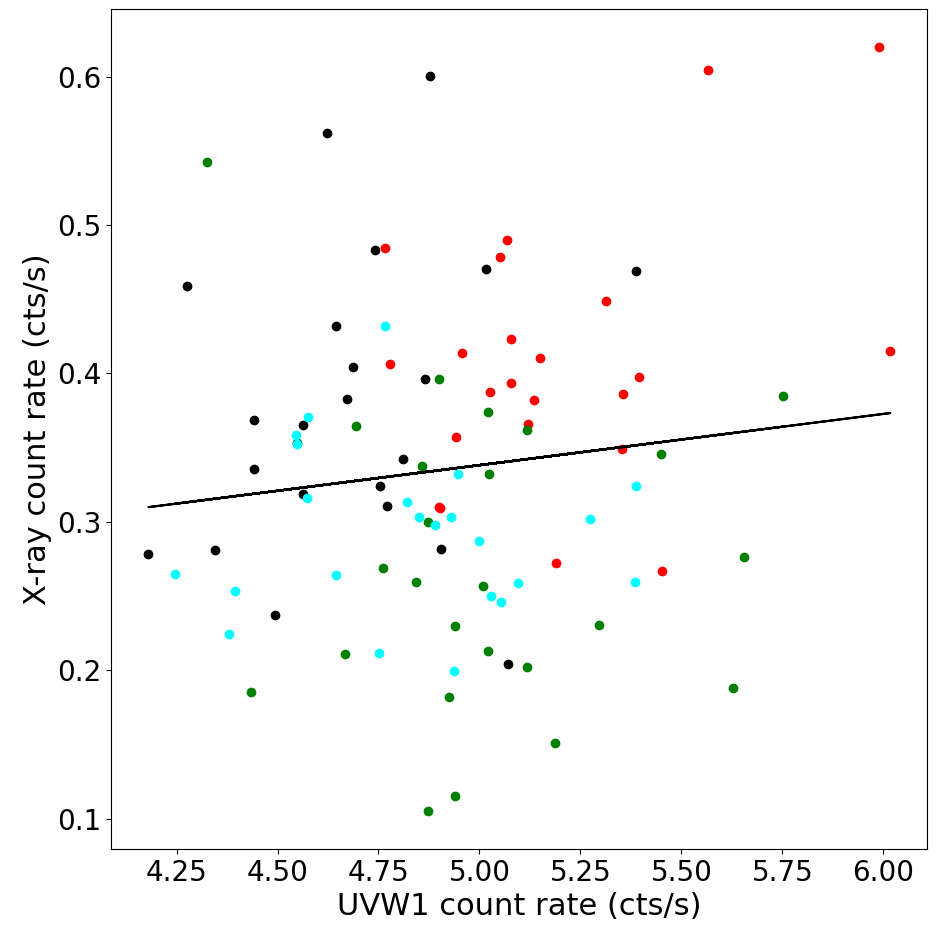}
	\includegraphics[width=\columnwidth, height=0.8\columnwidth]{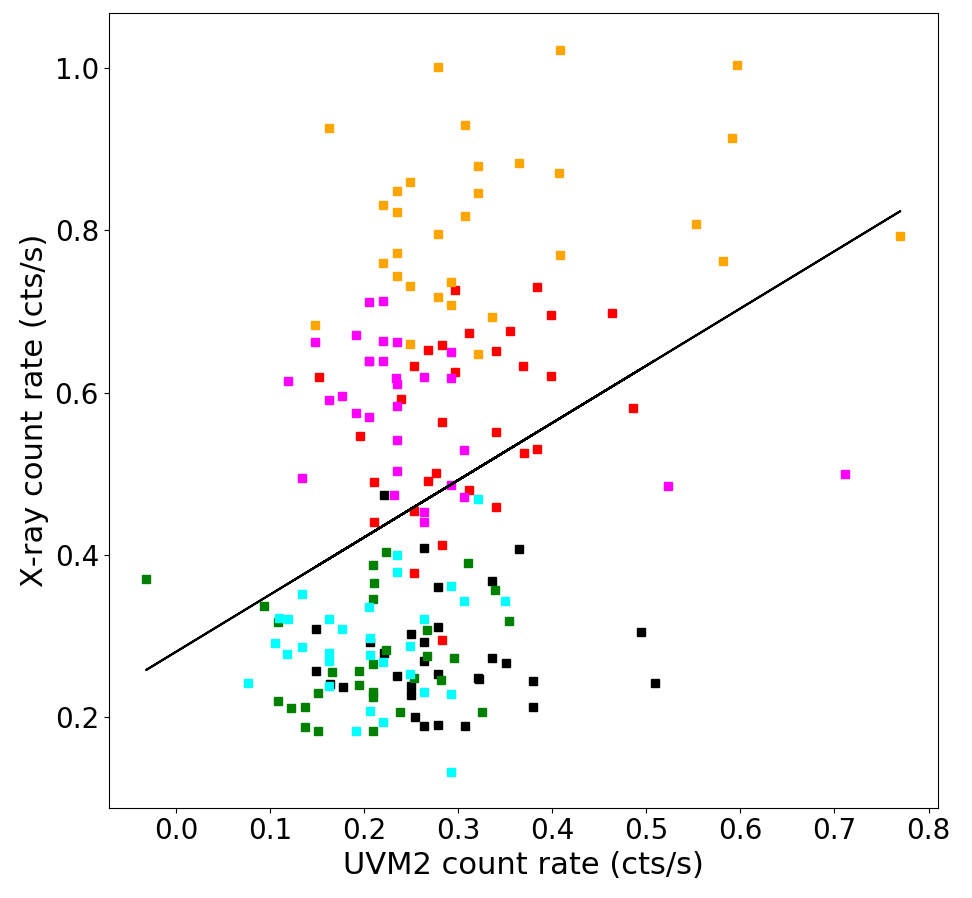}
	
    \caption{Comparison of X-ray and UV time resolved count rates. Top panel (A) The X-ray count-rate compared to the UVW1 count-rate when the lightcurves are binned in 100 second bins. The black line shows the negative correlation with a slope of -0.16. The colors indicate the 6 observations and are the same as in the spectral plots. Middle panel (B) shows that the anti-correlation is not present when data from the 4th and 5th observation are omitted.
    Lower panel (C) The same as panel A but comparing the X-ray to the shorter wavelength UVM2 countrate. In this case the data show a positive slope of 0.7  }
    \label{fig:X-ray_vs_UV_correlation}
\end{figure}

When binning the PDS 70 lightcurves into 2 ks time bins following the method of \cite{Schneider_2018} there is no evidence of any significant correlation between X-ray and UVW1. However, the observations of PDS 70 were much shorter, resulting in very few data points when binned at 2 ks. 
Further tests binning the lightcurve at 500 s and 100 s showed tentative evidence of the same anti-correlation. Fig.\ref{fig:X-ray_vs_UV_correlation} (A) shows that for 100 second time bins, the PDS 70 X-ray and UVW1 count-rates are slightly anti-correlated with a slope of -0.16 and a p-value of 0.00006 .
A p-value gives the probability that completely independent variables could have produced the same data, so this low p-value suggests that the anti-correlation, although small, is not just an artefact of this specific data set. The p-value was calculated using the Spearman's R test.


When comparing the X-ray countrate to the shorter wavelength UVM2 ligchturve in Fig.\ref{fig:X-ray_vs_UV_correlation}(C), a positive slope of 0.7 was found with a p-value of 0.00002
This is the opposite to what was found for the UVW1 data. The T-Tauri study did not include UVM2 data so it is not known if T-Tauri would show this difference between the UVW1 and UVM2 data. 

While this test is far from conclusive, it is interesting to note that PDS 70 appears to show the same X-ray - UVW1 anti-correlation as T-Tauri. However, it is not possible to conclude that coronal activity is being suppressed by enhanced accretion in PDS 70. The majority of the anti-correlation is caused by data points in Fig.\ref{fig:X-ray_vs_UV_correlation}(A) which are from the 4th (orange) and 5th (magenta) observation during the flare. It seems that the flare greatly enhanced the X-ray count-rate, and this happened to occur when the UVW1 count-rate was lower than average. The lack of a positive correlation does at least show that the X-ray and UVW1 are emitted by different regions or processes which appear be independent. Fig.\ref{fig:X-ray_vs_UV_correlation} (B) shows that there is no X-ray-UVW1 correlation, and a high p-value of 0.6, when data from the 4th and 5th observations are omitted. This supports the scenario where the X-ray and UVW1 emissions are from different processes during quiescence, revealing coronal and accretion activity respectively.

\cite{Ayres_2021} showed that when the X-rays are compared to chromospheric MgII 2800 \AA\ emission they always see a positive correlation for the Sun and several other non-accreting stars. For PDS 70 we see a negative correlation between UVW1 ($\sim$2900 \AA) which is the opposite to what is seen in the Sun and similar stars. 
However, at $\sim$2300 \AA\ (UVM2) we do see a positive correlation. 
UVW1 2900 \AA\ emission corresponds to a temperature of $\sim$8000 K which is more associated with the chromosphere. However, this is also the wavelength range where accretion emission would peak at $\sim$2800 \AA\ according to the model of \cite{Calvet_1998}. The accretion emission starts to drop off towards 2300 \AA. 
The UVM2 2300 \AA\ emission is from hotter 10,000 - 20,000 K plasma, so it includes a greater proportion of coronal plasma, and less from accretion emission or chromospheric emission.
So overall it seems there is a positive correlation between the X-ray and 2300 \AA\ emission which are both coronal, but an anti-correlation between the X-ray and 2800 \AA\ emission which are due to coronal and accretion emission respectively. In non-accreting stars, the X-ray and 2800 \AA\ emission do show a positive correlation. This suggests that for PDS 70 there is a process at work which produces excess 2800 \AA\ emission and is not closely linked to coronal processes. The most likely candidate is accretion, as shown by the accretion UV excess compared to expected photospheric emission in the model of \cite{Calvet_1998}.

\subsection{Comparison of PDS 70's X-ray properties with other T-Tauri stars}
\label{sec:Discussion_1_comparison_with_other_T-Tauris_AND_coronal} 

\begin{figure}
	\includegraphics[width=\columnwidth]{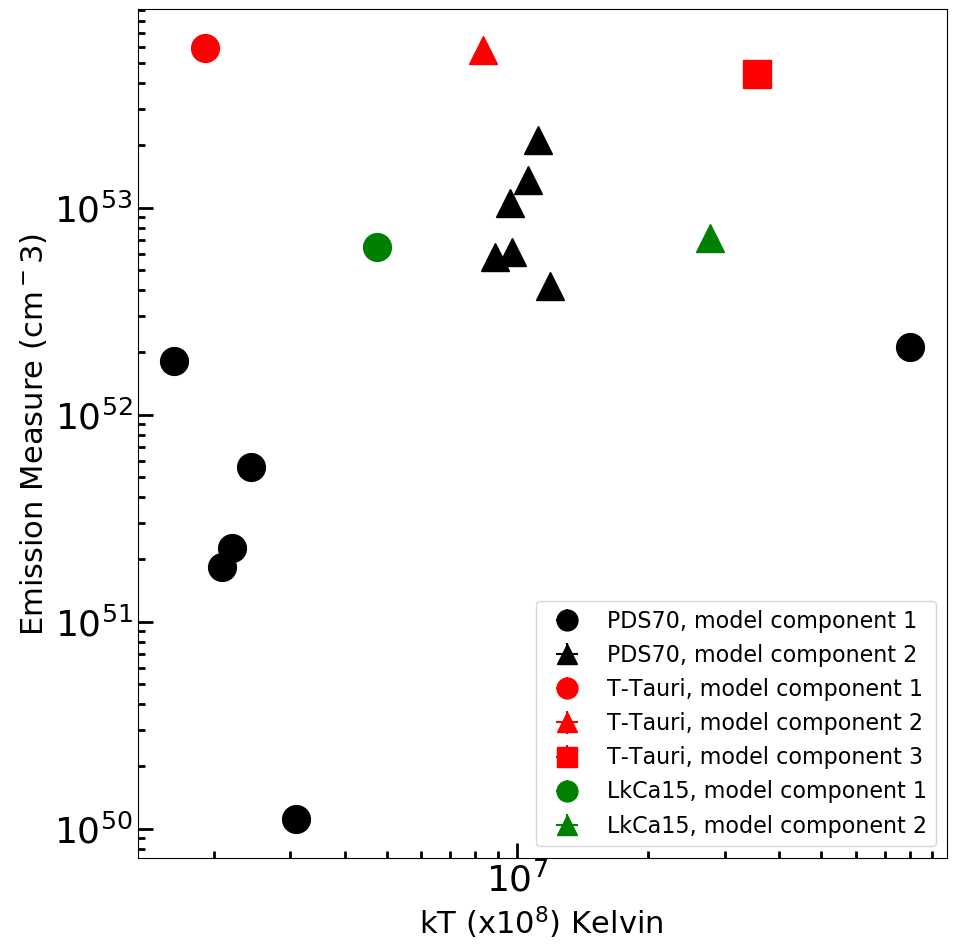}
    \caption{Comparing the EM and temperature from X-ray spectral fitting of PDS70 (black), LkCa15 (green) and T-Tauri (red). The spectra are fitted with plasma emission models which include two temperature components indicated by the circle and triangle sybols. The model for the T-Tauri spectrum (red) included a third component (square). The circle on the right shows the high temperature component required to fit the PDS 70 spectrum during the X-ray flare in the 4th observation.}
    \label{fig:PDS70_LkCA15_and_T_Tauri}
\end{figure}

\subsubsection{Comparison with the T-Tauri stars Lk Ca 15 and T-Tau}

Having fit the 6 spectra, comparisons can be drawn between the properties of the X-ray emission observed in PDS 70 and that observed from similar T-Tauri systems. Fig.\ref{fig:PDS70_LkCA15_and_T_Tauri} shows the fit parameters for all 6 PDS 70 spectra, compared to results from T-Tauri \citep{Schneider_2018} and LkCa 15 \citep{Skinner_2017}. T-Tauri has a high accretion rate of $4-8\times10^{-8}\ \mathrm{M_{\odot}/yr}$ compared to the estimated $\sim1\times10^{-10}\ \mathrm{M_{\odot}/yr}$ for PDS 70. LkCa 15 has a lower accretion rate of $\sim1\times10^{-9.2}\pm0.3\ \mathrm{M_{\odot}/yr}$  which is similar to the accretion rate of PDS 70 \citep{Donati_2019}. LkCa 15 and T-Tauri make a good comparison to PDS 70 because they have a similar spectral type, K5, K0 and K7 respectively. T-Tauri is included to provide a comparison with a higher accretion rate than PDS 70 or LkCa 15.

The emission measure (EM) is a function of the size of the emitting region and the electron density in the plasma (EM = $n^{2}_{e}V$). It is calculated from the normalisation of the VAPEC model using the equation(\ref{eqn:EM_from_norm}), where $d^{2}_{cm}$ is the distance to PDS 70 in cm.

\begin{equation}
\label{eqn:EM_from_norm}
\mathrm{EM} = 4\pi 10^{14}d^{2}_{\mathrm{cm}} \times \mathrm{normalisation}
\end{equation}

Also note that the temperature in Fig.\ref{fig:PDS70_LkCA15_and_T_Tauri} has been converted from units of keV to Kelvin. Fitting the X-ray spectra with several model components indicates the plasma temperature in each component, although in reality the plasma probably has a continuous range of temperatures. All 3 stars have a cool $\sim$ 2 Million Kelvin (MK) plasma component. In PDS 70 this makes a relatively small contribution to the X-ray spectrum, as shown by the low EM (black circles). The hotter component in PDS 70 is 10 MK, which is cooler than the $\sim$30 MK required for the hotter component in T-Tauri and LkCa 15. The notable exception is the temperature during observation 4 of PDS 70 where the temperature increased to 80 MK, as shown by the data point on the right.

The similar EM of PDS 70 and LkCa 15 shows that the emitting region is of a similar size, assuming a similar electron density. T-Tauri however has an order of magnitude larger EM, which requires a much larger emission region and/or increased electron density, which could indicate the contribution of an extended hot-spot, not present in the other two stars. 

\subsubsection{Comparison of PDS 70 with the $L_{\mathrm{X}}$ - $T_{\mathrm{av}}$ relation for weak-line T-Tauri stars}

\begin{figure}
	\includegraphics[width=\columnwidth]{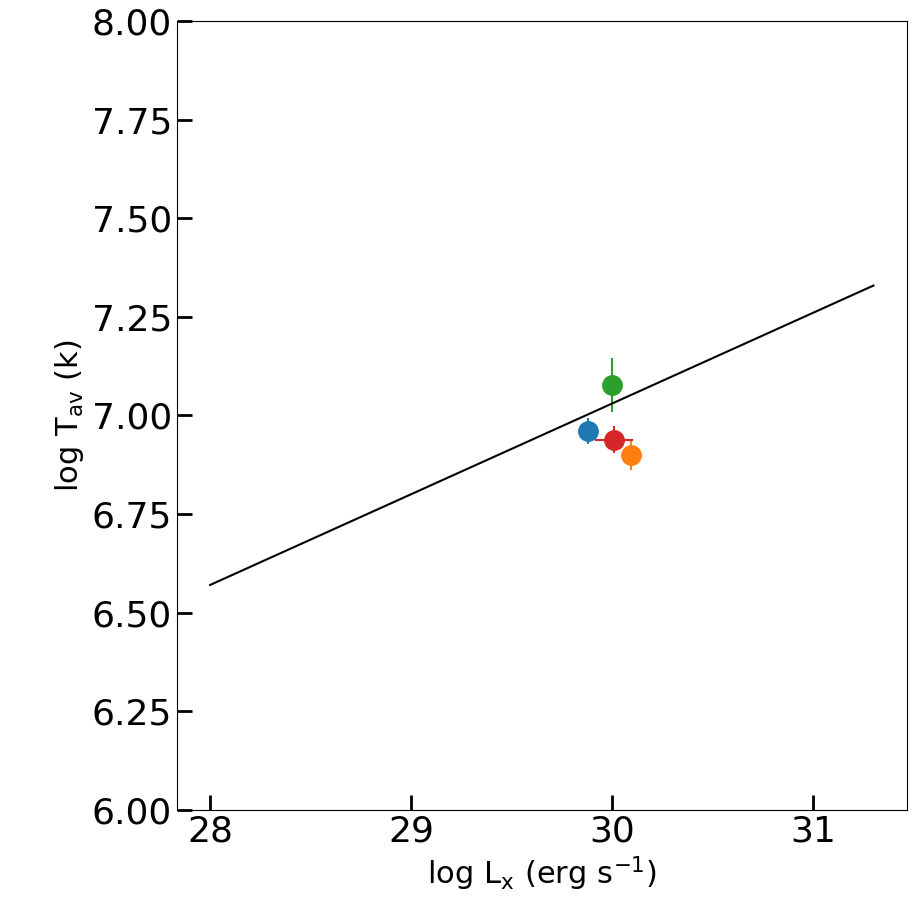}
    \caption{Comparing the electron temperature ($\mathrm{T_{av}}$) and X-ray luminosity ($L_{X}$) to the empirical relation for WTTS found in the sample of \citep{Telleschi_2007b}. The 4 data points correspond to the X-ray observations of PDS 70 which were not affected by flare activity (observations 1, 2, 3 and 6).  }
    \label{fig:PDS70_Lx_vs_Tav_WTTS}
\end{figure}

A correlation between between the X-ray luminosity and the coronal plasma electron temperature ($T_{\mathrm{av}}$) is seen in main-sequence stars \citep{Gudel_1997}. \cite{Telleschi_2007b} showed that non-accreting weak-line T-Tauris (WTTS) also follow a $L_{\mathrm{X}}$ - $T_{\mathrm{av}}$ relation. In contrast, their sample of accreting classical T-Tauris (CTTS) did not show any $L_{\mathrm{X}}$ - $T_{\mathrm{av}}$ correlation. According to their linear regression fits, the $L_{\mathrm{X}}$ and $T_{\mathrm{av}}$ parameters for the CTTS sample had a 43- 80 per cent probability of being uncorrelated, whereas the WTTS had a < 0.01 per cent probability of no correlation. The linear regression fit to the WTTS sample resulted in an empirical relation 
$\mathrm{log}\ T_{\mathrm{av}} = (0.23\pm0.03)\ \mathrm{log}\ L_{\mathrm{X}} +(0.13\pm0.93)$. Although the process underlying the $L_{\mathrm{X}}$ - $T_{\mathrm{av}}$ relation is not clearly established, \cite{Telleschi_2007b} concluded that it is the presence of accretion which disrupts this relation for CTTS.

With the rate of accretion in PDS 70 being very low, we tested whether its $L_{\mathrm{X}}$ and $T_{\mathrm{av}}$ values are still consistent with the empirical $L_{\mathrm{X}}$ - $T_{\mathrm{av}}$ relation. The $L_{\mathrm{X}}$ values are the unabsorbed values in the 0.2 - 12 keV range given in Table.\ref{Table:Table_3_spec_fit_results}. The average temperature is calculated from the 2-temperature fit parameters using a weighted mean, where the contribution is weighted according to the EM of each model component. 

We find that during the non-flaring periods, the $L_{\mathrm{X}}$ and $T_{\mathrm{av}}$ properties of PDS 70 are indeed consistent with the empirical relation. Fig.\ref{fig:PDS70_Lx_vs_Tav_WTTS} shows the PDS 70 values from observation 1, 2, 3 and 6 compared to the \cite{Telleschi_2007b} relation (straight line). This indicates that the accretion rate of $\sim 10^{-10}\ \mathrm{M_{\odot}\ yr^{-1}}$ (see section \ref{sec:Discussion_1_Accretion}) is low enough that it does not disrupt the process underlying the $L_{\mathrm{X}}$ - $T_{\mathrm{av}}$ relation seen in non-accreting WTTS and main-sequence stars.



\subsubsection{Comparison of PDS 70 X-ray luminosity with that expected from coronal emission}

The coronal X-ray luminosity of stars is related to their rotation period and spectral type, due to convection zone depth, by well established empirical relations (e.g. \citealt{Pallavicini_1981, Johnstone_2021}). These relations show that when X-ray luminosity is scaled to the size of the star by taking the $\mathrm{L_{X}/L_{bol}}$ ratio, the maximum 'saturated' luminosity is $\mathrm{log\ L_{X}/L_{bol}=}$ -3.1 Stars remain at this saturated luminosity until the rotation declines below a certain rate dependent on spectral type. After this, the activity declines with decreasing rotation rate as the star spins down over time. If the X-ray emission of PDS 70 is coronal in origin, it would be expected to be consistent with the X-ray - rotation relation, and by extension, the X-ray - age relation.

The quiescent unabsorbed $\mathrm{L_{X}}$ of $1\times10^{30}$ erg s$^{-1}$ found for PDS 70 is at the level expected for a very young 0.6 - 0.8 $\mathrm{M_{\odot}}$ star according to the models of \cite{Johnstone_2021}. In Table.\ref{Table:Table_3_spec_fit_results} the unabsorbed $\mathrm{L_{X}/L_{bol}}$ is calculated for each observation using the bolometric luminosity of 0.35 $\mathrm{L_{bol}}$ for PDS 70 \citep{Pecaut_Mamajek_2016}. The ratio for each of the quiescent observations is close to the saturation limit log -3.1 This is as expected for a young rapidly rotating star. 

PDS 70 has a 3.03 day rotation period which is shorter than the 3.1 day limit \citep{Pizzolato_2003} for a 0.79 $\mathrm{M_{\odot}}$ star to be exhibiting saturated X-ray emission. For the 4th and 5th observation, PDS 70 temporarily exceeded the saturation limit. 

We conclude that the X-ray emission of PDS 70 is consistent with that expected for coronal emission for a young, rapidly rotating star emitting at the saturation limit.

\subsubsection{The X-ray flare properties}

The increased X-ray emission during the 4th and 5th observation is potentially due to a coronal flare. To investigate if this is the cause, it is useful to compare this event to expected flare properties seen for similar stars. The following comparison is based on flares detected in a Chandra observation of 28 PMS stars in the Orion Nebula Cluster \citep{Wolk_2005}. The sample is similar to PDS 70 in that they are spectral type K5-K7 and ages 0.5 - 19 Myr. The comparison sample are slightly higher mass than PDS 70, ranging from 0.9 – 1.2 M$_{\odot}$.
When studying the flare properties of the PMS stars, \cite{Wolk_2005} distinguish between the characteristic (non-flaring) X-ray flux, periods of elevated flux and finally flares. The distinguishing feature of flares is that their morphology shows an initial sharp rise to a peak luminosity, followed by an exponential decay. 

For PDS 70, the countrate is already high at the beginning of the 4th observation, so the sharp rise, if it occurred, was not observed. There is also uncertainty about the decay phase as the 5th observation appears to show this, but only if it is assumed that the X-ray flux declined smoothly in the gaps between observations.  With these caveats in mind, the PDS 70 flare can be compared to the flares in the ONC sample by looking at characteristics such as luminosity, duration and frequency of flares.

The peak luminosity $\mathrm{L_{X}}$ averaged over the 14ks 4th observation of PDS 70 was $2.82\times10^{30}$ erg/s.
The PMS sample had a peak $\mathrm{L_{X}}$ range of $5\times10^{29}$ to $5\times10^{31}$ erg/s. Direct comparisons are difficult because flare amplitudes are distributed as a decreasing power-law in occurrence rate, with frequent small flares and infrequent large flares.  So although there is no 'average' flare property, PDS 70 is within the expected range for flare $\mathrm{L_{X}}$ seen on other K-type PMS stars. 

For the PMS sample, flare durations range from 1 h to 3 days  (3.6 ks to 260 ks), with the most common duration being 80 ks. The duration of the PDS 70 flare is uncertain as the complete flare was not observed. As an upper limit, the maximum duration could be $\sim$172 ks if the flare started just after observation 3. The minimum is $\sim$102 ks if it began just as observation 4 started.
This is within the expected range of 3.6 - 260 ks observed for K-type PMS stars in Orion, but is longer than the average flare duration of 80 ks.

The ratio of flaring $\mathrm{L_{X}}$ to non-flaring $\mathrm{L_{X}}$ covered a wide range for the PMS sample, with the median ratio being 3.5x and the maximum observed in one case being 27x.
For PDS 70 the flare $\mathrm{L_{X}}$ of $2.82\times10^{30}$ and the average non-flaring $\mathrm{L_{X}}$ of $1\times10^{30}$ gives a ratio of 2.8x. By this measure, the PDS 70 event was well within the increase in luminosity seen on other similar stars. 

The PMS sample exhibited on average 1 flare per star every 7.5 days (650 ks). The PDS 70 total observation was 6 days long and 1 flare was detected. However, the significance of this comparison is limited as the number of detected flares is strongly dependent on detection limits. There may be many more flares which are not bright enough to be detected depending on the stellar distance.

Overall, the PDS 70 X-ray flare has properties consistent with flares seen in a sample of PMS K-type stars \citep{Wolk_2005}. In terms of $\mathrm{L_{X}}$, duration, frequency and ratio of peak $\mathrm{L_{X}}$ to non-flaring $\mathrm{L_{X}}$, the PDS 70 flare is within the expected ranges for a K-type PMS star and does not exhibit any properties which would suggest that it is not caused by normal coronal processes.


\subsection{Accretion}
\label{sec:Discussion_1_Accretion} 

\begin{figure}
	\includegraphics[width=\columnwidth]{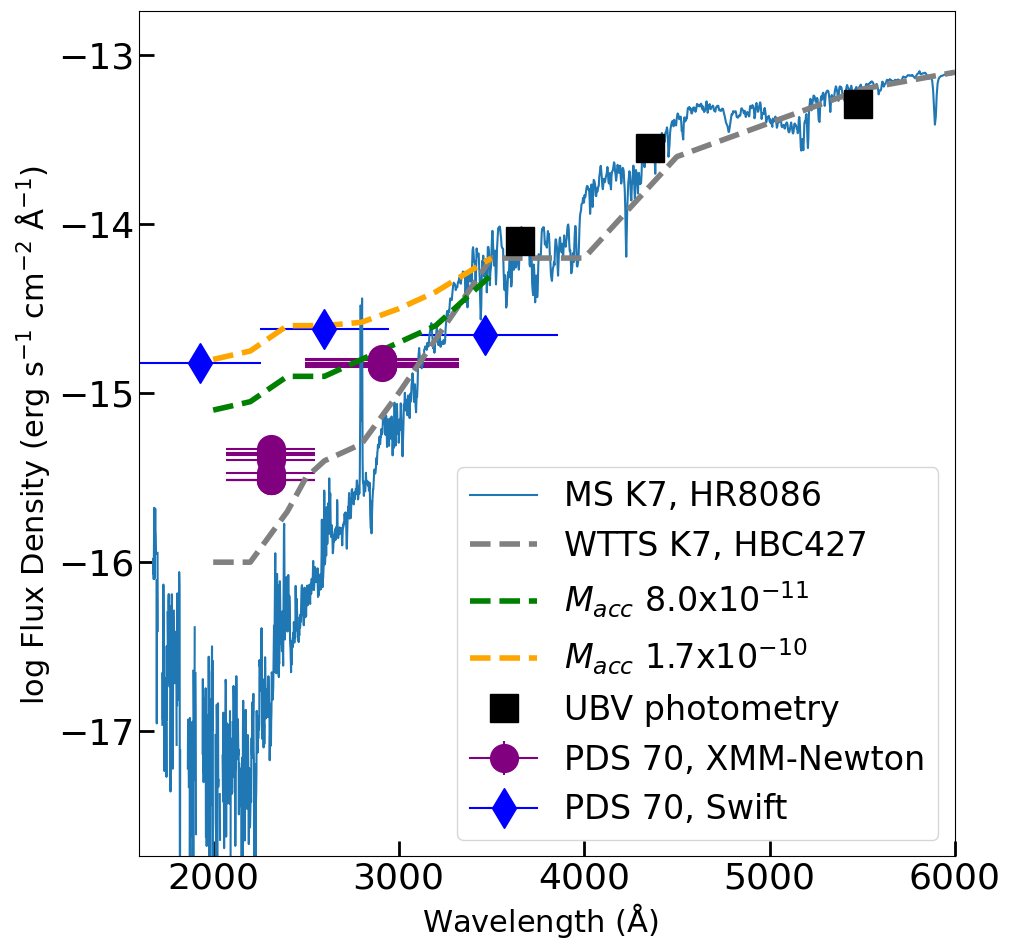}
    \caption{UV flux density of PDS 70 from XMM-OM data (purple circles) and Swift-UVOT observations 1 year earlier (blue diamonds). Also plotted are UBV photometry from Gregorio Hetem et al. 2002 (black squares). The spectrum is a non-accreting main-sequence K7 type star HR8086 from the HST STIS Next Generation Spectral Library and has been re-scaled to the distance and stellar radius of PDS70. The excess UV emission detected for PDS 70 at wavelengths $<3200$ \AA, is attributed to accretion emission. The grey dashed line indicates the spectrum for a K7 type WTTS (HBC 427) (Ingleby et al. 2013) which shows additional active chromospheric emission compared to the main sequence star, but still less UV excess than PDS 70. The green and orange lines indicate the additional UV emission expected for different accretion rates.}
    \label{fig:UV_flux_density_with_K7_HR8086_comparison_spec}
\end{figure}

Accretion of disc material onto the star produces excess UV emission compared to the coronal activity. In Fig.\ref{fig:UV_flux_density_with_K7_HR8086_comparison_spec} the UV flux density measurements from Table.\ref{Table:Table_3_spec_fit_results} are compared to the expected emission for a K7 type star. The comparison spectrum is from a real star so it includes UV emission from the corona, transition region, chromosphere and photosphere. The template spectrum is the non-accreting main-sequence K7 type star HR8086 available from the STIS Next Generation Spectral Library. This has been rescaled to account for the difference in stellar radius and distance. Also plotted (grey dashed line) is the flux for the K7 WTTS star HBC427 from \cite{Ingleby_2013}. It is known that young stars have more active chromospheres and this contributes to the UV excess in the 2000 - 3200 \AA\ range when compared to the main-sequence star.

The flux from PDS 70 shows an excess at wavelengths shorter than 3200 \AA, where the emission from a stellar corona declines rapidly. The analysis by \cite{Thanathibodee_2020} showed that the \textit{Swift} 2019 data is consistent with the emission expected for a low accretion rate of $1.7\times 10^{-10}\ \mathrm{M_{\odot}\ yr^{-1}}$ based on the accretion shock model of \cite{Calvet_1998} added to the WTTS spectrum. We compare the UV flux density measurements from the XMM-OM in 2 filters (purple circles) with the measurements from a year earlier taken by \textit{Swift} (blue diamonds). The results are broadly consistent, although there is a slight decline in the UV emission detected by \textit{XMM} during July 2020, which could indicate a reduction in the accretion rate, lower than the $8\times 10^{-11}\ \mathrm{M_{\odot}\ yr^{-1}}$ rate modelled in \cite{Thanathibodee_2020}. UV spectra observed using \textit{HST} in December 2020 measured a UV continuum flux consistent with that seen in the \textit{XMM} photometry, and also estimated an accretion rate of $2.6 (0.5-14.8)\times 10^{-10}\ \mathrm{M_{\odot}\ yr^{-1}}$ based on the luminosity of the C$_{\mathrm{IV}}$ line \citep{Skinner_2022}. The variability seen in the XMM UV photometry suggests that the accretion rate may be fluctuating on timescales of days to months, and this could explain some of the differences in estimated accretion rates from different studies.


\subsection{XUV effect on disc evolution}
\label{sec:Discussion_1_disc_evolution} 

\begin{figure}
	\includegraphics[width=\columnwidth]{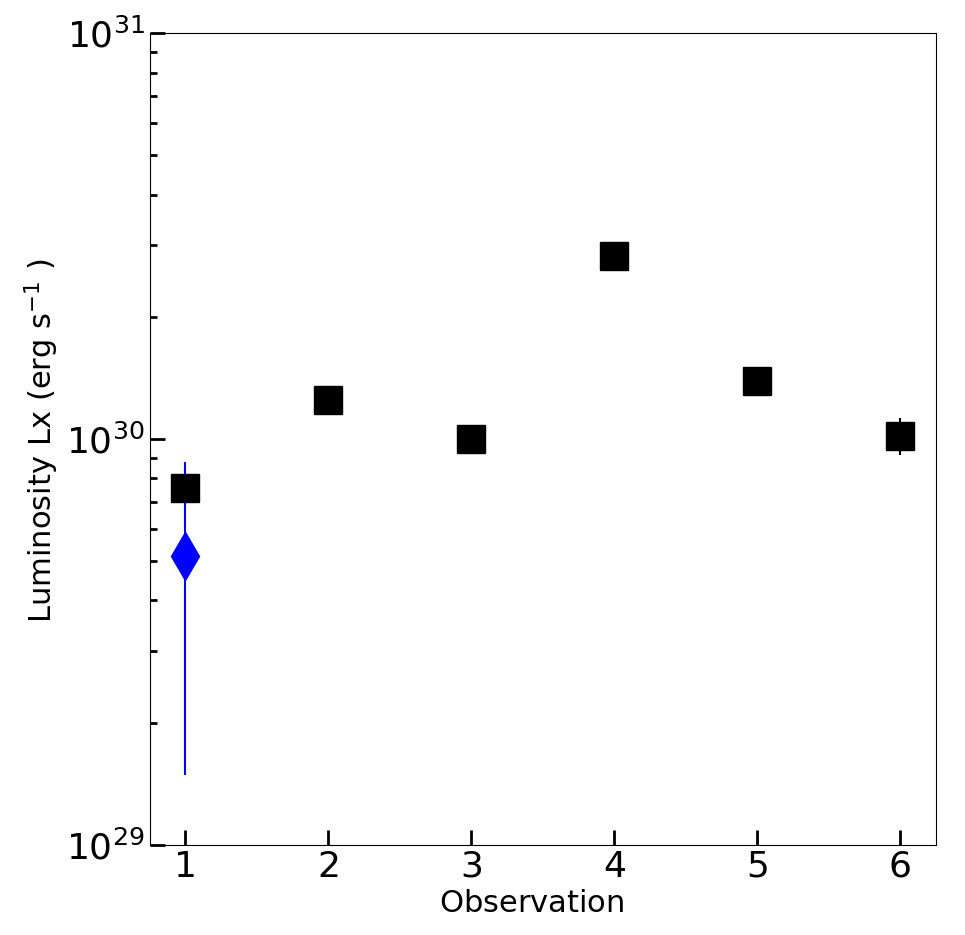}
    \caption{X-ray luminosity from the 6 XMM observations in 2020 (black squares) and the Swift observation (blue diamond) from 2019. Error bars on the XMM data points are smaller than the markers. }
    \label{fig:PDS70_Lx_XMM_vs_Swift}
\end{figure}

The X-ray luminosity of PDS 70 for each of the 6 observations is shown in Fig.\ref{fig:PDS70_Lx_XMM_vs_Swift}. The X-ray luminosity detected by Swift is also shown (blue diamond) and is consistent with the $\sim 1 \times 10^{30}\ \mathrm{erg\ s^{-1}}$ detected during most of the \textit{XMM} observations (black squares). This is further evidence that this represents the normal X-ray emission state of PDS 70, while the 4th and 5th observations represent a temporarily enhanced emission. During this episode, the $\mathrm{L_{X}}$ increased by a factor of $\sim$3, from $\sim 1 \times 10^{30}\ \mathrm{erg\ s^{-1}}$, to over $\sim 2.8 \times 10^{30}\ \mathrm{erg\ s^{-1}}$. 

The original analysis based on the \textit{Swift} $\mathrm{L_{X}}$ estimated that the evaporation rate of the protoplanetary disc is $\dot{M}_{\mathrm w}\simeq 10^{-8}$\,M$_{\odot}$\,yr$^{-1}$ which would result in total dispersal of the currently remaining disk within $M_{\mathrm d}/\dot{M}_{\mathrm w}<1$Myr \citep{Joyce_2020} for standard assumptions (i.e., a gas-to-dust ratio of 100) and a total disc mass $M_{\mathrm d} \lesssim 10^{-2}$--$10^{-3}$M$_{\odot}$ (e.g. \citealt{Keppler_2018}). The confirmation of the $\mathrm{L_{X}}$ detected by \textit{Swift} with multiple \textit{XMM} spectra of much greater S/N confirms that this is the appropriate $\mathrm{L_{X}}$ to use for estimating the disc evaporation rate. 



\section{Conclusions}

1) We confirm the detection of PDS 70 as an X-ray and UV source.

2) The X-ray luminosity is confirmed to be $\mathrm{L_{X, unabsorbed}} =$  $1\times10^{30}$ erg s$^{-1}$ and this is likely to be the normal quiescent luminosity as it was detected in the majority of observations over 6 days, and matched the previous \textit{Swift} measurement.

3) The UV data show emission in excess of the expected coronal emission for a K7 type star. This supports the finding of \cite{Thanathibodee_2020} that accretion is still taking place from the inner disk onto the star at a rate of approximately $1\times 10^{-10}\ \mathrm{M_{\odot}\ yr^{-1}}$. However, we find that the UV excess, and hence the accretion rate, might have been slightly reduced during the \textit{XMM} observations compared to the \textit{Swift} observations 1 year previously and also compared to the \textit{HST} observations 5 months later \citep{Skinner_2022}. This could indicate some variability in the accretion over months.

4) A large increase in the X-ray count rate was detected during one observation, which was matched by an increase in the plasma temperature 
as determined from spectral fitting with a 2-temperature optically-thin plasma model. 

5) The increased X-ray flux was not detected during the same phase of the 1st rotation of the star. This indicates that this increased flux was due to a transient event rather than different areas of the stellar surface coming into view. Comparison of this event with X-ray flares seen on other K-type T-Tauri stars showed that it was consistent with the duration, energy and other characteristics of these flares, and we conclude that this event is most likely a coronal flare rather than being associated with accretion.

6) This study confirms the previous finding \citep{Joyce_2020}, that with the present level of X-ray irradiation, and reasonable estimates for the remaining mass of the protoplanetary disc, final dispersal will be complete in $\sim$ 1 Myr. This will bring an end to the current phase of planet formation. 
This would mean the total protoplanetary disc lifetime of PDS 70 will be $\sim$ 6 Myr based on its current age estimate.

7) The X-ray luminosity and plasma electron temperature measured from PDS 70s quiescent X-ray spectra are consistent with the $L_{\mathrm{X}}$ - $T_{\mathrm{av}}$ relation seen in non-accreting weak-line T-Tauri stars and main-sequence stars, indicating that the level of accretion in PDS 70 is not sufficient to disrupt this relation, as it does in CTTS.


\section*{Acknowledgements}

We thank the anonymous reviewer for their helpful comments which improved the manuscript.
This work was carried out as part of the ExoplANETS-A project http://exoplanet-atmosphere.eu (Exoplanet Atmosphere New Emission Transmission Spectra Analysis);
https://cordis.europa.eu/project/rcn/212911 en.html; The ExoplANETS-A project was funded from the EU's Horizon-2020 programme; Grant Agreement no. 776403. RA acknowledges funding from the European Research Council (ERC) under the European Union's Horizon 2020 research and innovation program (grant agreement No 681601).

\section*{Data Availability}

The data used in this study are available from the XMM-Newton Science archive at \textit{http://nxsa.esac.esa.int/nxsa-web}. The observation numbers are 0863801201 to 0863801701 and were observed for the proposal "Disc dispersal in action: XUV observations of the T-Tauri star PDS 70", PI Joyce.



\bibliographystyle{mnras}
\bibliography{example} 









\appendix

\section{Source detection in X-ray and UV}

\begin{figure}
	\includegraphics[width=\columnwidth]{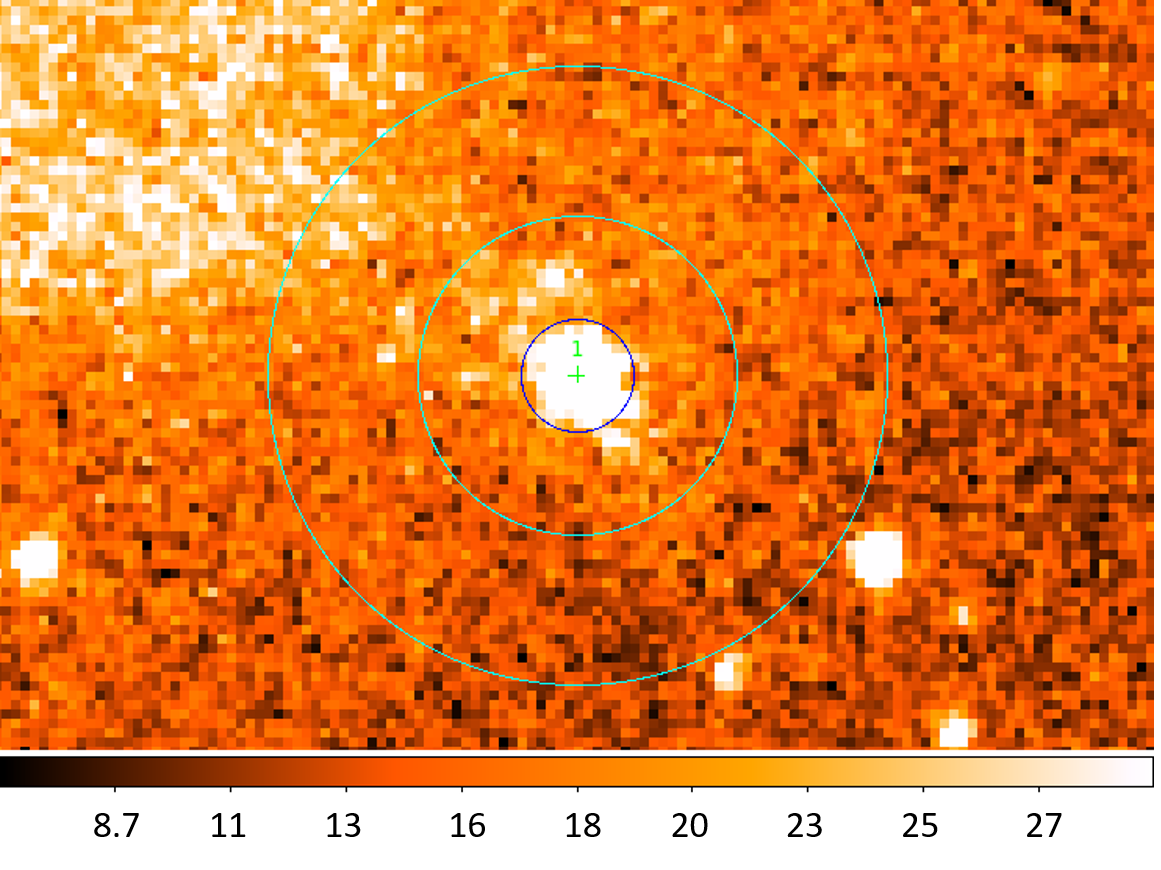}
	\includegraphics[width=\columnwidth]{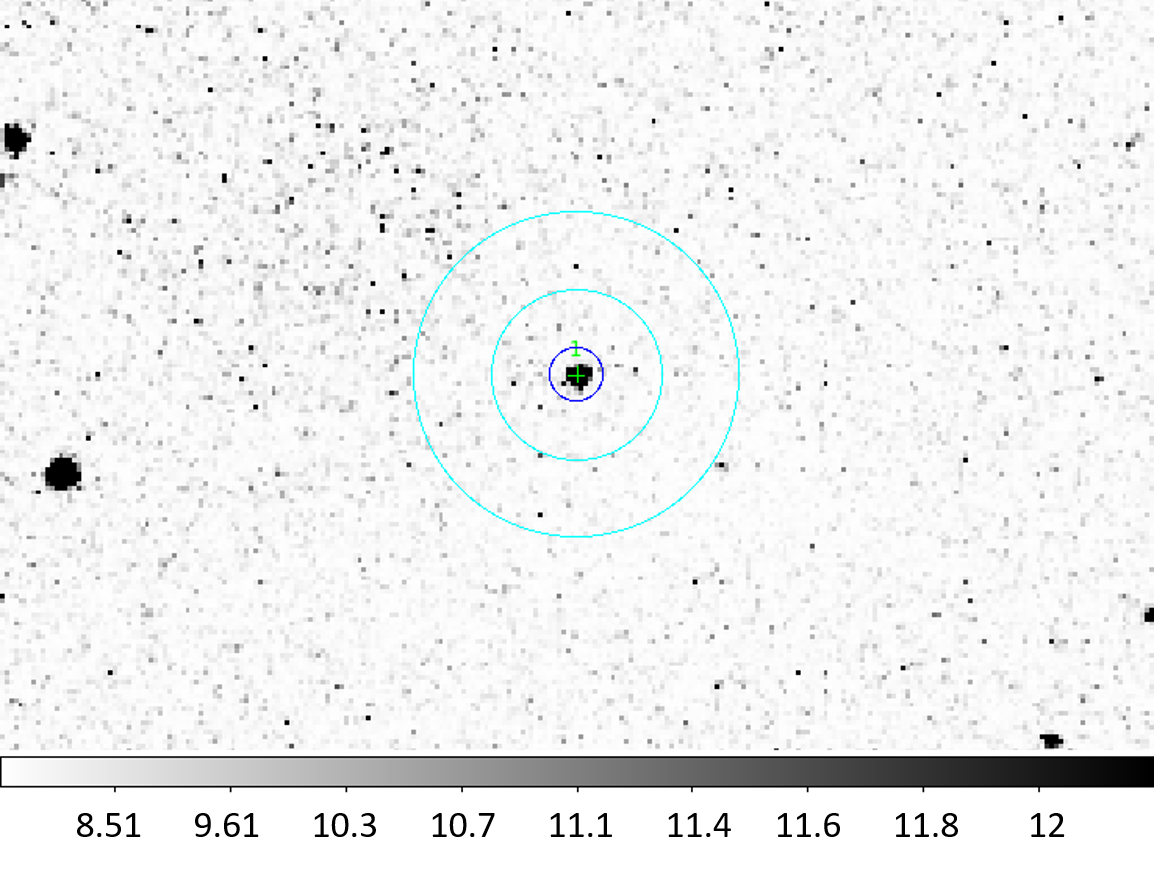}
    \caption{UV source detection. Top panel (A), source detection in the OM-UVW1 (2495-3325 \AA) filter. Lower panel (B), detection in the OM-UVM2 (2070-2550 \AA) filter. The source counts are extracted from the inner circle, and the background is measured from the outer annulus. The colours indicate total counts in each pixel according to the scale on the x axis. }
    \label{fig:Fig_Appendix1_XUV_source_detection_image}
\end{figure}
\begin{figure}
	\includegraphics[width=\columnwidth]{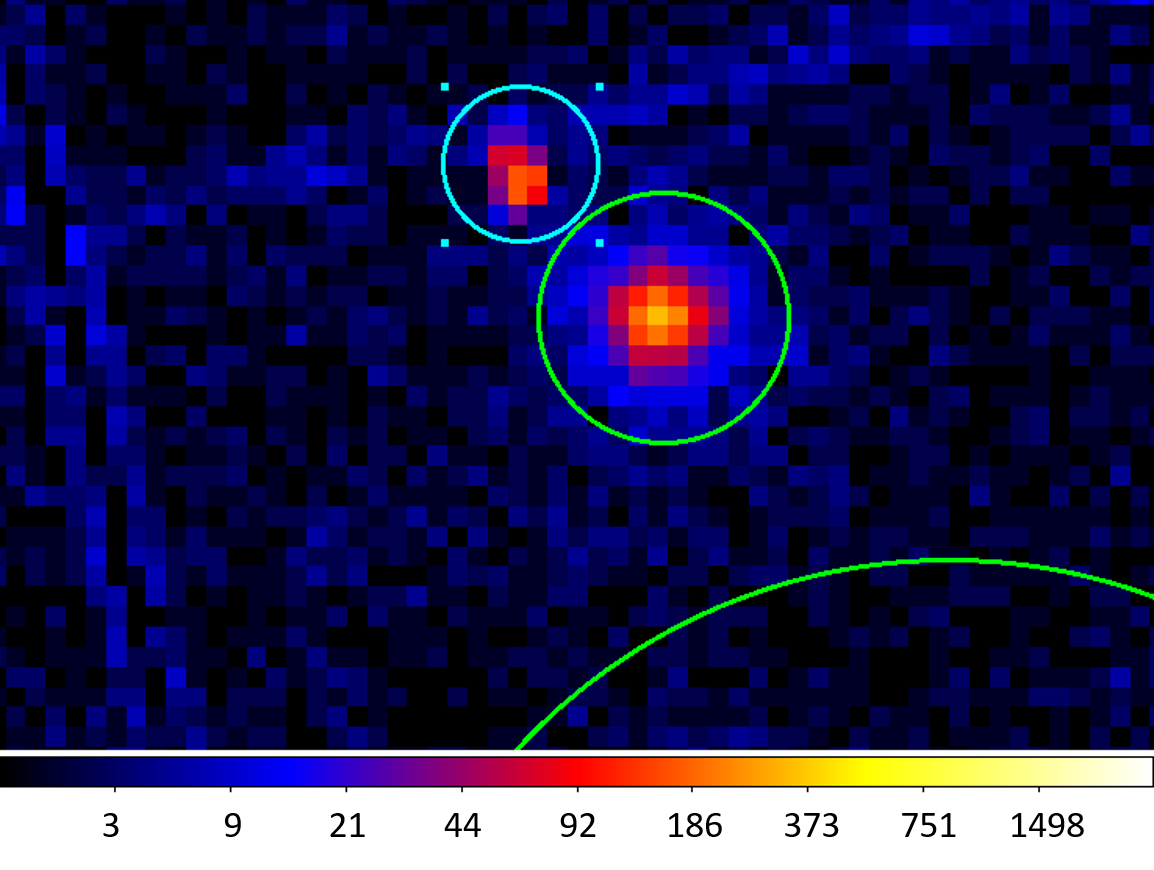}
    \caption{Source detection for the 2nd observation with the PN X-ray camera showing the transient source in the smaller circle at RA=14:08:12.54, Dec=-41:23:27.85  The transient source is not detected in the MOS1 or MOS2 image.}
    \label{fig:Fig_Appendix2_transient_source_image}
\end{figure}

\begin{figure}
	\includegraphics[width=\columnwidth]{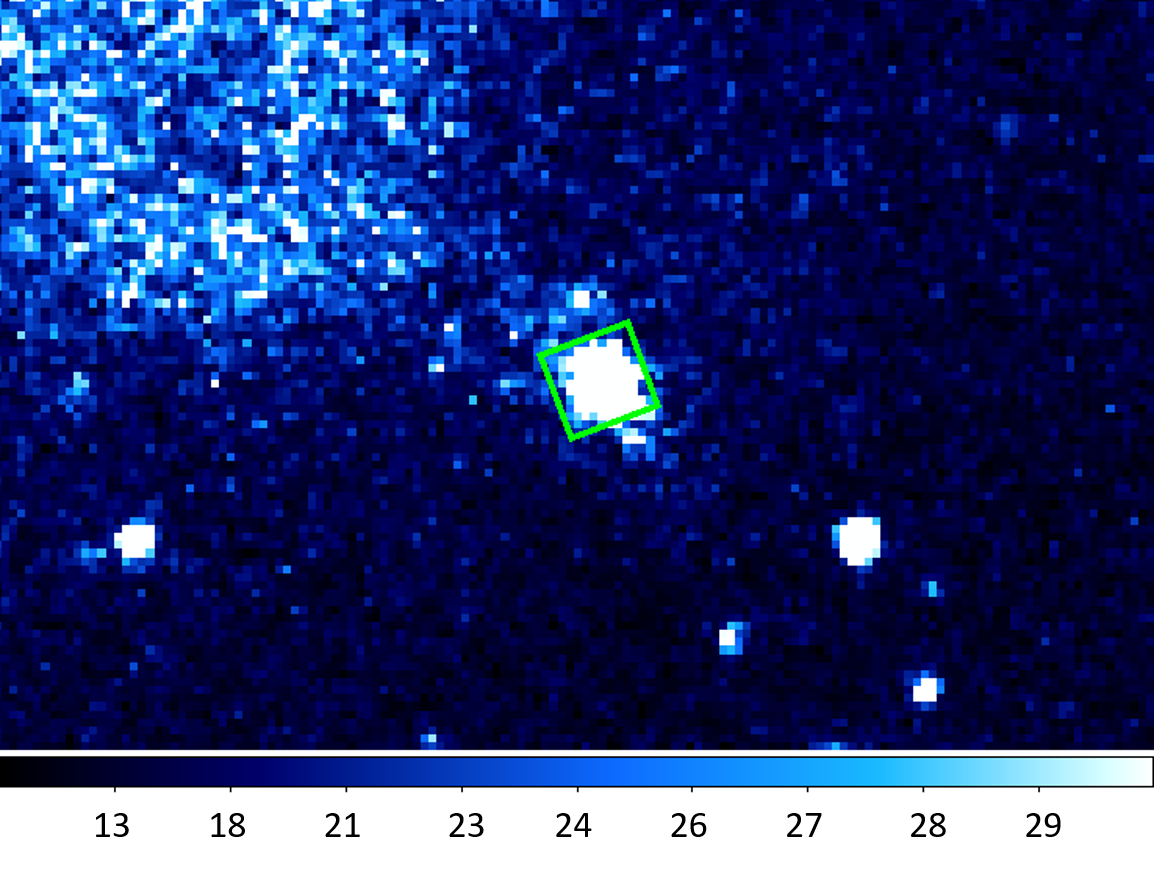}
    \caption{Optical Monitor UVW1 filter image showing the position of the fast-mode window used to extract UV light-curve data.}
    \label{fig:Appen_3_UVW1_show_full_psf_and_FAST_MODE_WINDOW}
\end{figure}

\begin{figure}
	\includegraphics[width=\columnwidth, height=75mm]{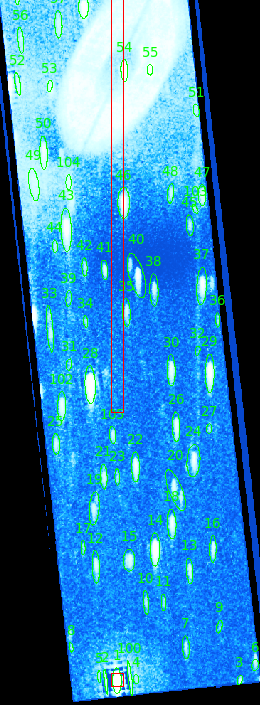}
    \caption{UV Grism 2-D spectral image showing the target detected at 0th order (red square at the bottom of the image), which includes optical light, and the extraction region for the 1st order UV spectrum (red rectangle). No significant UV spectrum is detected. The sources which produce lines in the pipeline processing products 1-D spectra are not attributed to PDS 70.}
    \label{fig:Appen_4_Fig_example_UV_Grism_2D}
\end{figure}


\bsp	
\label{lastpage}
\end{document}